\documentclass[12pt,preprint]{aastex}
\slugcomment{Revised Version: \today}

\usepackage{color}
\usepackage{graphicx}

\newcommand{\dij}{}

\newcommand{\nnhp}{N$_2$H$^+$}
\newcommand{\nhhh}{NH$_3$}
\newcommand{\nn}{N$_2$}
\newcommand{\co}{CO}
\newcommand{\ceo}{C$^{18}$O}
\newcommand{\hh}{H$_2$}

\newcommand{\mum}{$\,\mu$m}

\begin{document}
\title{Dense Gas Tracers in Perseus: Relating the N$_2$H$^+$, NH$_3$, and Dust Continuum Properties of Pre- and Proto-Stellar Cores.}
\author{
        Doug Johnstone\altaffilmark{1,2},
        Erik Rosolowsky\altaffilmark{3},        
        Mario Tafalla\altaffilmark{4},
        \& Helen Kirk\altaffilmark{1,2}
}
\altaffiltext{1}{National Research Council Canada, Herzberg
Institute of Astrophysics, 5071 West Saanich Rd, Victoria, BC, V9E
2E7, Canada; doug.johnstone@nrc-cnrc.gc.ca}
\altaffiltext{2}{Department of Physics \& Astronomy, University of Victoria,
Victoria, BC, V8P 1A1, Canada}
\altaffiltext{3}{University of British Columbia Okanagan, 
Kelowna, BC, V1V 1V7, Canada}
\altaffiltext{4}{ Observatorio Astron\'omico Nacional (IGN), Alfonso XII 3, 
E-28014 Madrid, Spain}

\begin{abstract}
We investigate 35 pre-stellar cores and 36 proto-stellar cores in the Perseus molecular cloud.
We find a very tight correlation between the physical parameters describing the \nnhp\ and \nhhh\ gas. Both the velocity centroids and the line widths of \nnhp\ and \nhhh\ correlate much better than either species correlates with \co, as expected if the nitrogen-bearing species are probing primarily the dense core gas where the \co\ has been depleted. We also find a tight correlation in the inferred abundance ratio between \nnhp\ and para-\nhhh\ across all cores, with $N$(p-\nhhh)/$N$(\nnhp)$= 22 \pm 10$. We find a mild correlation between \nhhh\ (and \nnhp) column density and the (sub)millimeter dust continuum derived \hh\ column density for pre-stellar cores, $N$(p-\nhhh)/$N$(\hh) $\sim 10^{-8}$, but {\it do not} find a fixed ratio for proto-stellar cores. 

The observations suggest that in the Perseus molecular cloud the formation and destruction mechanisms for the two nitrogen-bearing species are similar, regardless of the physical conditions in the dense core gas. While the equivalence of \nnhp\ and \nhhh\ as powerful tracers of dense gas is validated, the lack of correspondence between these species and the (sub)millimeter dust continuum observations for proto-stellar cores is disconcerting and presently unexplained.
\end{abstract}


\section{Introduction}
\label{sec:intro}

The study of dense cores in nearby
clouds like Taurus and Perseus has 
provided  a main observational
input for testing theories of star formation.
These cores are currently forming solar-type stars, and the
study of their internal structure offers a unique opportunity
to explore observationally the initial conditions
of stellar birth and the first stages of proto-stellar evolution.

The systematic improvement in sensitivity and angular resolution 
of IR and radio observations has gradually revealed
that dense cores have a more complex internal structure than
initially thought. Studies of the density in pre-stellar cores, 
for example,
have found radial profiles with a central flattening reminiscent
of the isothermal (Bonnor-Ebert) models, although the overall
state of equilibrium of these cores is still a matter of debate
\citep[e.g.,][] {dif07,war07}.
The chemical composition of the core material also presents 
systematic variations with radius. The innermost gas in a core prior 
to star formation is usually depleted of C-bearing molecules
like CO, CS, or HCO$^+$, which are thought to freeze out on the
cold ($\approx 10$~K) dust grains at gas densities of a few
$10^4$ cm$^{-3}$  \citep[][]{cas99,ber02,taf02}.
This molecular freeze out affects 
our ability to determine the physical conditions and kinematics of
the pre-stellar gas, as it limits the number of tracers available to
observers and adds a layer of complexity when comparing
the emission from different molecular species. Because of this,
understanding and characterizing molecular freeze out has 
become a necessary step to realize the potential of core studies
\citep[e.g.,][]{ber07}.

Most of our information on molecular freeze out comes from
the detailed study of a selected group of cores and globules,
like L1544, B68, L1498, and L1517B (see previous references).
These works have shown that under freeze out conditions, the
most reliable tracers of the dense core material are nitrogen-bearing 
molecules like \nnhp\ and \nhhh\ together with the dust component. 
These three tracers systematically provide consistent views of
the dense cores, in terms of their maps having similar
shape, size, and  peak position, and their molecular
spectra presenting similar central velocity and linewidth.
Such a good agreement between tracers suggests that their emission 
arises from the same gas in a core, and that their chemical
similarities dominate over a number of significant differences 
between their emission properties. The $J,K$=(1,1) transition of
\nhhh, for example, has a rather low critical density of
$\sim 10^3\,$cm$^{-3}$  \citep{eva99},
while the commonly-used $J$=1--0 transition of \nnhp\ has a much higher value 
\citep[$\sim 10^5\,$cm$^{-3}$,][]{ung97}, and the dust emission is
directly proportional to the gas column density but sensitive to
the dust temperature and emissivity \citep{hil83}.

Although the main observational characteristics of molecular 
freeze out seem 
well established by now, a number of unsolved issues require 
further investigation. Most previous work, for example, used
reduced and highly selected samples of targets, so detailed
radiative transfer modeling could be carried out \citep[e.g.,][]{cas99, ber02, taf04}. Such a 
strong target selection, unfortunately
limits the statistical significance of the work and precludes 
the investigation of topics like cloud-wide chemical
variations, time evolution of the core composition, and a
critical inter-comparison between the behavior of the 
freeze out resistant
tracers (\nnhp, \nhhh, dust continuum). To investigate these
issues, it is necessary to carry out systematic, multitracer observations 
of a large sample of cores in a cloud, 
something that was not possible just a few years ago.
Fortunately, a number of surveys of nearby molecular clouds 
have been recently undertaken across a wide range of wavelengths and molecular
species \citep[see for example,][]{rid06}, and these have provided
unique data sets of continuum emission and molecular line diagnostics.

One star-forming region which has
received significant recent attention is the Perseus molecular cloud
\citep[e.g.,][]{hat05, eno06, jor06, kir06,  kir07, reb07, ros08, hat09}. This
cloud contains
clustered low and intermediate mass proto-stellar candidates,
and seems to represent a case in
between the low-mass star-forming cloud of
Taurus and the more massive Orion star-forming region. A distance of
$250\pm50\,$pc to Perseus has been adopted by the {\it Spitzer} c2d
team \citep{eva03}, following measurements by \citet{cer93} and
\citet{bel02}. \citet{jor07, jor08}  combined the Spitzer observations
and the submillimeter continuum data of this cloud to produce a
complete sample of deeply embedded protostars in Perseus and to
determine the clustering properties of the pre-stellar cores and
protostars. From a combined \nnhp\ and \ceo\ survey of this Perseus core
population, \citet{kir07} showed 
that the gas motions in the vicinity of submillimeter cores vary significantly, 
from relatively quiescent inside the individual cores to
dynamic in the surrounding gas. Further observations of the Perseus core
population have been recently presented by 
\citet{ros08}. These authors \citep[also][]{fos09,sch09}  
have used \nhhh\ and CCS observations to study,
among other parameters, the gas kinetic temperature in the cores 
and its variation between isolated and clustered environments.

The above data set of Perseus core observations provides
a unique opportunity to carry out a statistical,
cloud-wide comparison between
\nnhp, \nhhh, and the dust continuum, the three 
most reliable
tracers of the dense gas in cores. To this end, we have combined 
all available observations of these tracers in a
large set of
both pre-stellar and proto-stellar cores, and we have
carried out a new analysis of the data deriving 
excitation and column densities in an
homogeneous manner (checked later against
detailed radiative transfer modeling). As a result of this
work, we present here the first statistically significant
comparison between the so-far believed robust tracers
of the star-forming core material, and show that while
the tracers do indeed present similar behavior,
significant deviations occur in the densest,
most evolved core population.

This paper is organized in the following manner. Section~\ref{sec:obs}
presents the observational data used in the analysis.
Section~\ref{sec:obscomp} compares the physical properties derived
from the observed \nnhp\ and \nhhh\ spectra. The chemical properties
of the cores are discussed in \S\,\ref{sec:abund}. 
All of the observations are placed in context with core evolution in \S\,\ref{sec:disc}. The major
conclusions of the paper are summarized in \S\,\ref{sec:concs}.

\section{A Coordinated Observational Data Set for Perseus}
\label{sec:obs}

As mentioned in the Introduction, the Perseus molecular cloud has been amply surveyed with {\it Spitzer\ } in the mid-infrared \citep[][]{jor06, reb07} and with ground-based telescopes at (sub)millimeter wavelength dust emission \cite[][]{hat05, eno06, kir06}. At the locations of pre-stellar and proto-stellar cores, the Perseus molecular cloud has also been observed in \nnhp\ (1--0) and \ceo\ (2--1)  \citep[][]{kir07} and \nhhh\ (1,1) \citep[][]{ros08}. As well, near infrared extinction mapping using 2MASS sources \citep[for the methodology, see][]{lom01} has been used to provide for the large-scale structure of the cloud \citep[][]{kir06}.

\subsection{Submillimeter and Mid-Infrared Observations}
\label{sec:obs:cont}

\citet{kir06} analyzed a 3.5 degree$^2$ region of the Perseus molecular cloud using 850\mum\ data taken with SCUBA at the JCMT. At this wavelength, the beam size of the observations is about 
15\arcsec, although smoothing of the map resulted in an effective beam of $\sim$20\arcsec. They identified 58 submillimeter cores and found that a majority of them could be well fit by stable Bonnor-Ebert spheres \citep[][]{ebe55, bon56}. Comparing the locations of the submillimeter cores with the underlying column density in the cloud measured via near infrared extinction, \citet{kir06} concluded that the cores are preferentially found in the highest column density regions of the molecular cloud. These zones have relatively large mean densities as well, $\left< n\right>  > 5\times10^3\,$cm$^{-3}$, as derived from the extinction maps \citep[see Table 3 in][]{kir06}. The majority of the molecular cloud mass, however, was found to exist at low column density arguing that only a small fraction of the cloud is participating in the star formation process. 

A further comparison between the 850\mum\ data and {\it Spitzer\ } mid-infrared observations was performed by \citet{jor07}. In this survey, 72 submillimeter cores were identified (the slight change in number between the two surveys being due to the resolution of the reconstructed submillimeter map and the clump-finding thresholds used to identify objects), of which half were identified as harboring protostars. As expected, the submillimeter cores coincident with protostars were found on average to be brighter (more massive) and more centrally peaked in appearance. Also, when discernible, the protostars were found to be centrally located within the cores.

A similar map of the Perseus molecular cloud was obtained by \citet{eno06} at 1.1\,mm using Bolocam at the Caltech Submillimeter Observatory. At this wavelength the effective beam size of the observations is about 30\arcsec. The 7.5 degree$^2$ region was inspected and 122 cores were identified. In regions of overlap with the 850\mum\ SCUBA map, the two core catalogues are very similar. {\dij Due to its larger beam size,} the 1.1\,mm Bolocam data, is sensitive to somewhat more extended {\dij low-surface brightness} sources.

\subsection{\nnhp\ and \co\ Observations}
\label{sec:obs:nnhp}

Using the IRAM 30-meter telescope, \citet{kir07} simultaneously observed \ceo\ (2--1) and \nnhp\ (1--0) toward 150 candidate dense cores in the Perseus molecular cloud. At these transitions, the effective beam sizes are about 11\arcsec\ and 25\arcsec, respectively. For the 89 sources selected by 850\mum\ emission, 84\,\% yielded detectable \nnhp\ emission and all were observable in \ceo. The hyperfine structure of the \nnhp\ (1--0) emission was utilized to fit for the physical properties of the dense gas, including the velocity of the line centroid, the line width, the excitation temperature, and the line optical depth. For the \ceo\ (2--1) observations, the line centroid and line width were measured.

\citet{kir07} found that the dense gas associated with the \nnhp\ pointings displays nearly thermal line widths, particularly for the subset that appear starless \citep[as determined by][]{jor07}. This result is consistent with other surveys of dense gas, which used \nhhh\ \citep[][]{ben89, jij99}, and reinforces the notion that cores are supported primarily by thermal pressure, large non-thermal motions having disappeared on small scales ($< 0.1\,$pc) and at high densities ($> 10^4\,$cm$^{-3}$). On the other hand, the \ceo\ (2--1) observations revealed that the lower density, un-depleted gas retains significant non-thermal motion. Interestingly, the offset between the velocity centroid of the \ceo\ emission and the velocity centroid of the \nnhp\ emission was found to be less than the sound speed for 90\,\% of the targets, arguing that the two regions of emission are nevertheless coupled (see also \S~\ref{sec:obscomp:cent}).

\subsection{\nhhh\ Observations}
\label{sec:obs:nhhh}

Using the GBT, \citet{ros08} observed \nhhh\  (1,1) and (2,2) toward 193 dense core candidates in the Perseus molecular cloud, drawn primarily from the 1.1\,mm and 850\mum\ continuum surveys. For the observed \nhhh\ transitions, the effective beam size is about 30\arcsec. Ammonia emission was found toward nearly all submillimeter sources and the hyperfine structure of the \nhhh\  lines were fit for the observational properties of the emitting region, including the velocity of the line centroid, the line width, the excitation temperature, and the line optical depths. As well, through comparison of the \nhhh\  (1,1) and (2,2) observations, the rotation and kinetic temperature of the gas and the non-thermal contribution to the line width was determined. For the cores in Perseus, a typical low kinetic temperature of $T_k = 11\pm 2\,$K was measured \citep{ros08}. As found for the \nnhp\ observations \citep{kir07}, the \nhhh\ lines  are usually quite narrow and thermally-dominated.

\subsection{Correlating the Data Sets}
\label{sec:obs:corr}

Despite the coordinated approach to data taking inside the Perseus molecular cloud, the individual pointing observations in \nnhp\ and \nhhh\ were not explicitly aimed at the same locations on the sky. The \nnhp\ observations were taken primarily toward 850\mum\ SCUBA locations, and a subset of `by eye' extinction locations observed on digitized POSS-II Palomar plates. The \nhhh\ observations typically were  taken toward peaks in the Bolocam map.  In general, however, the deviation between the two pointings was significantly less than 25\arcsec\ (within the \nnhp and \nhhh\ beams).

For the purpose of this study, we collected all observations in \nnhp\ and \nhhh\ which were less than 25\arcsec\ apart and include them in Table~\ref{tab1}. {\dij Only ten of these sources have offsets larger than 15\arcsec\ and the mean offset is  9\arcsec.} Table~\ref{tab1} lists the source names from the studies by \citet{kir07} and \citet{ros08}, as well as the location of the source, and the offset distance between the two surveys. Also presented in Table~\ref{tab1} are the integrated line intensities for \nhhh\ (1,1), \nnhp\ (1--0), and \ceo\ (2--1), as well as the (sub)millimeter flux from SCUBA (850\mum) and Bolocam (1.1\,mm). {\dij The uncertainties in the line intensities are typically less then ten percent, while the uncertainty in the (sub)millimeter flux is about twenty percent. The (sub)millimeter fluxes are measured at the location of the \nhhh\ sources and averaged over 30\arcsec}.

Table~\ref{tab1} lists 82 individual cores. Of these 82 cores, three have no detection in \nnhp\ (all of these are also found to be weak in \nhhh). An additional seven cores have poor parameter fits for \nnhp, due to their low optical depth and the tight covariance between excitation temperature and optical depth in such a regime (see Appendix~\ref{sec:app}). One additional core could not be fit using the \nhhh\ lines. Thus, of the 82 cores obtained, 71 contain enough information to be useful in the detailed comparison of their physical  properties. Of these, 35 are pre-stellar cores and 36 are proto-stellar cores, according to the analysis of \citet{jor07}.  For these 71 sources, Table~\ref{tab2} lists the physical properties derived from fitting to the hyperfine components \citep[see][]{kir07, ros08}. 

\section{Physical Properties of the Perseus Cores}
\label{sec:obscomp}

In this section we compare the observationally derived physical parameters for the \nnhp\ (1--0) and the \nhhh\ (1,1) molecular line transitions. It is worth reminding the reader that the effective beam sizes of the two measurements, at their associated telescopes, are 25\arcsec\ and 30\arcsec, respectively. As well, the formal critical densities for thermalizing the emission are $\sim 10^5\,$cm$^{-3}$ and $\sim 10^3\,$cm$^{-3}$, respectively\footnote{Note, however, that the critical density is not an exact indicator of the conditions under which emission is most efficient. This is especially true for low frequency transitions, such as \nhhh\ (1,1), where stimulated emission from the Cosmic Background has a substantial effect on the detailed balance of the energy levels.}. Thus, the ammonia, with its lower critical density and somewhat larger beam, should probe to lower density gas if it is present in appreciable quantities.

\subsection{Comparison of Centroid Velocities}
\label{sec:obscomp:cent}

\begin{figure} 
\epsscale{0.8}
\includegraphics[width=1.0\textwidth]{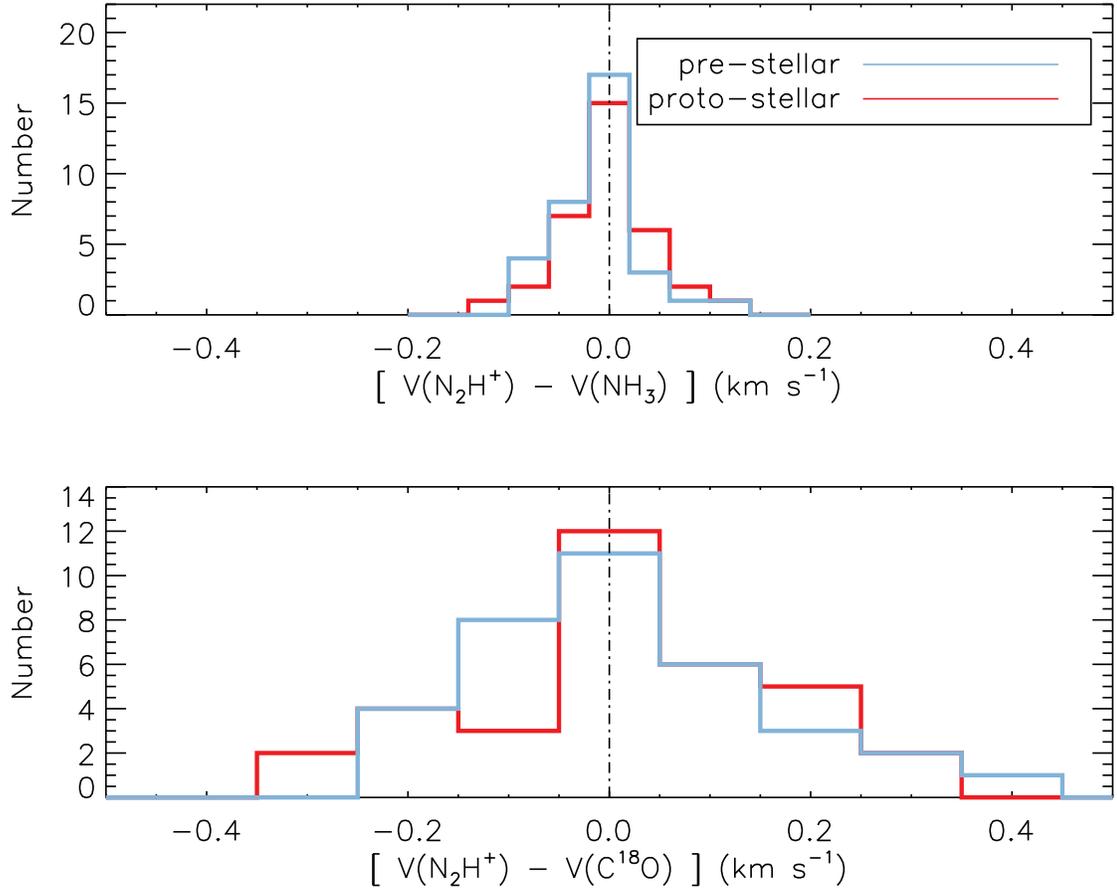}
\caption{Histograms of the variation in the centroid velocity between the different chemical species for the observed cores in Perseus. In both panels the blue lines indicate pre-stellar cores while the red lines indicate proto-stellar cores. The top panel compares \nnhp\ observations with \nhhh, while the bottom panel compares \nnhp\ with \ceo\ (see text for details).}
\label{fig:centroid}
\end{figure}

Figure~\ref{fig:centroid} plots a histogram of the offset in centroid
velocities between the \nnhp\  and \nhhh\ (upper panel) and between
the \nnhp\ and \ceo\ (lower panel). 
{\dij Ten percent of the \nnhp\ pointings were found to have multiple velocity components \citep{kir07} while
considerably fewer of the \nhhh\ observations needed multiple fits \citep{ros08}. In this paper, we
consider only the closest velocity match between the nitrogen-bearing species (and \ceo) at each position.} 
In this figure (and all subsequent
plots), red denotes proto-stellar sources while blue denotes
pre-stellar sources. Despite the fact that the \nnhp\ and \ceo\
observations were taken simultaneously, toward the exact same location
on the sky, the \nnhp\ and \nhhh\ centroid agreement is significantly
stronger. Indeed, the mean absolute offset is only 0.07\,km\,s$^{-1}$,
much smaller than the sound speed in the gas ($c_s \sim
0.20\,$km\,s$^{-1}$ for $T_k = 11\,$K) and about the accuracy of the
\nnhp(1--0) line rest frequency \citep{pag09}.
As noted by
\citet{kir07}, the mean absolute offset between the \nnhp\ and
\ceo,~0.14\,km\,s$^{-1}$ for pre-stellar cores and 0.17\,km\,s$^{-1}$
for proto-stellar cores, is of order the sound speed in the gas  and,
interestingly, much smaller than the typical \ceo\ (2--1) non-thermal
line widths \citep[see \S~\ref{sec:obscomp:width} or][]{kir07}.

If the \nnhp\ and \nhhh\ emission is coming from within the dense core while the \ceo\ emission is produced on larger scales in the material surrounding the dense core, then the strong correlation between the centroid velocities of the nitrogen-bearing molecules is expected.  The critical density for \nhhh\ (1,1) is, however, similar to that required to excite the \ceo\ (2--1) line and thus one might have expected a contamination of the core \nhhh\ (1,1) measurement by the surrounding \ceo-rich cloud. Unlike Taurus, where the mean density in the cloud is  low and only the cores reach densities greater than a few $\times 10^3\,$cm$^{-3}$, in Perseus the high extinction zones in which the (sub)millimeter cores are found have significant density, $\left< n\right>  > 5\times 10^3\,$cm$^{-3}$, as derived from the extinction maps \citep[see Table 3 in][]{kir06}.

It thus appears that the \nhhh\ (1,1) line is not significantly contaminated by the bulk material surrounding the cores in Perseus and that both the observed \nhhh\ (1,1) and \nnhp\ (1--0) emission is produced within the dense cores.

\subsection{Comparison of Line Widths}
\label{sec:obscomp:width}

\begin{figure} 
\epsscale{0.8}
\includegraphics[width=0.9\textwidth]{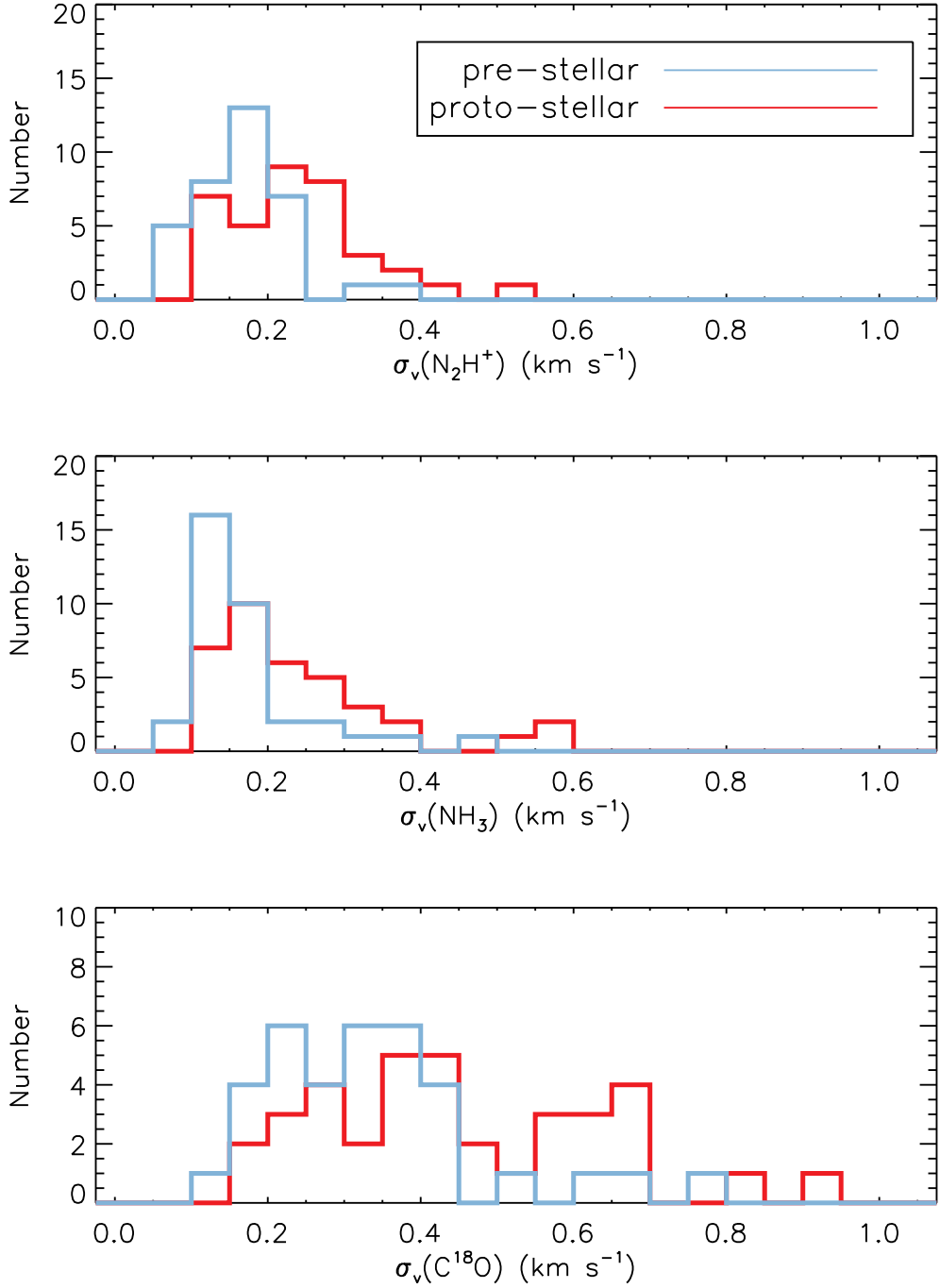}
\caption{Histograms of the measured line widths (in units of Gaussian $\sigma$) for the observed cores in Perseus. In all panels the blue lines indicate pre-stellar cores while the red lines indicate proto-stellar cores. The top through bottom panels show the results for \nnhp, \nhhh, and \ceo, respectively.}
\label{fig:sigma}
\end{figure}

As a second test of the environments in which the various lines are produced, Figure~\ref{fig:sigma} shows histograms of the observed line widths (in units of Gaussian $\sigma$), uncorrected for thermal broadening,  for \nnhp\ (upper panel), \nhhh\ (middle panel), and \ceo\ (lower panel). As noted by both \citet{kir07} and \citet{ros08}, the \nnhp\ and \nhhh\ line widths do not display significant non-thermal motions. This is especially true for the measured line widths of the pre-stellar sources (blue histograms), where very few measurements fall beyond the sound speed ($c_s \sim 0.2\,$km\,s$^{-1}$). The \ceo\ (2--1) line exhibits a quite different behavior. In most cases, both pre-stellar and proto-stellar, the measured line width is dominated by non-thermal motions. 

As with the line centroid histogram, it is striking how the \nnhp\ and the \nhhh\ histogram measurements agree, in contrast to the comparison with \ceo\ observations.  Despite the similar critical densities for \nhhh\ and \ceo, there is no hint of a correlation between these two measurements. Again, it would appear that the \nhhh\ (1,1) line is not significantly contaminated by the bulk material surrounding the dense cores. 

\subsection{Comparison of Physical Properties}
\label{sec:obscomp:tex}

When determining the column densities and abundances of \nnhp\ and \nhhh\ in the next section, it is necessary to utilize the observationally fit physical properties to the hyperfine structure of each molecule's observed transition. In the preceding sections, the line centroid and line width have been investigated. This section compares the key fitting parameters source by source.

\begin{figure} 
\epsscale{0.8}
\includegraphics[width=1.0\textwidth]{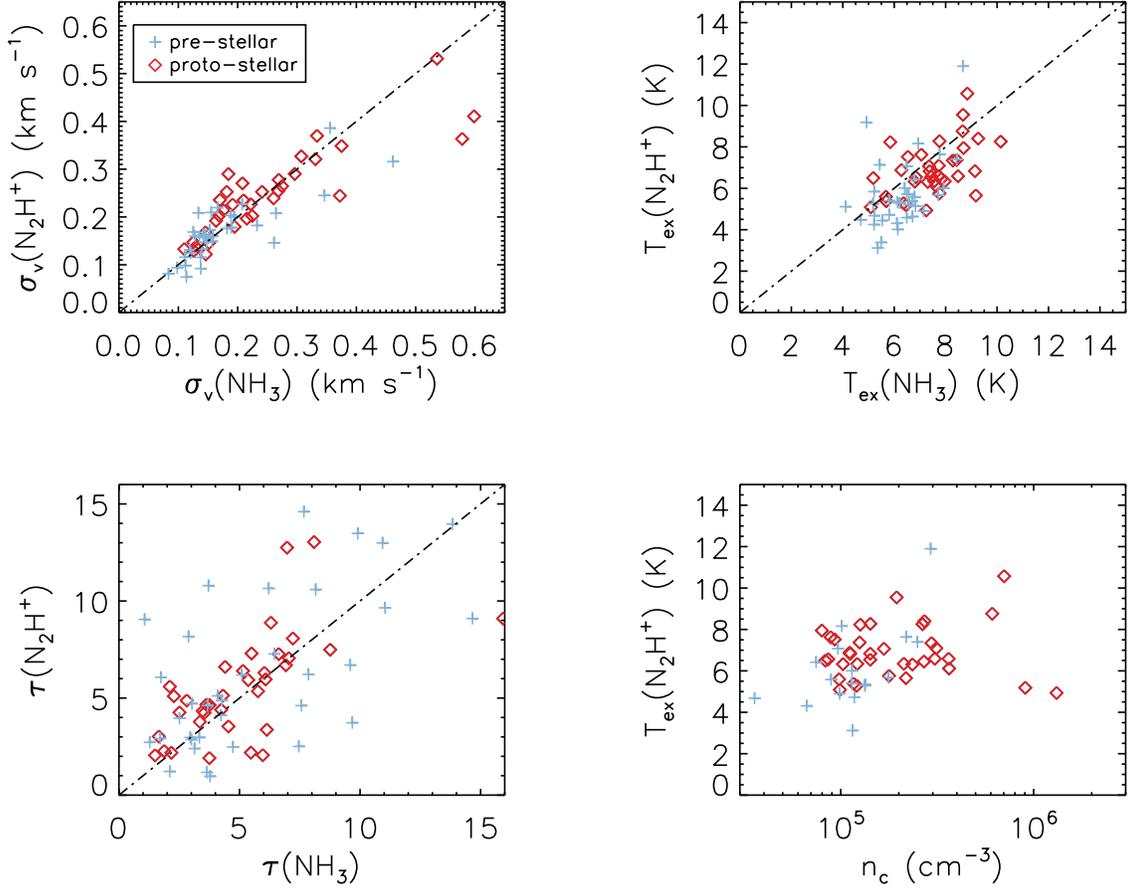}
\caption{Correlations among the physical properties fit to the hyperfine components of \nnhp\ and \nhhh\ for the observed cores in Perseus. In all panels the blue plus signs indicate pre-stellar cores while the red diamonds indicate proto-stellar cores. The dash-dotted lines denotes a one-to-one correspondence. The top left panel plots the measured line widths of \nnhp\ against \nhhh. The top right panel plots the derived excitation temperature, $T_{\rm ex}$, of \nnhp\ against \nhhh. The bottom left panel plots the derived optical depth, $\tau$, of \nnhp\ against \nhhh. The bottom right panel plots the derived $T_{\rm ex}$ of \nnhp\ against the estimated density in the core, 
$n_c$ (see text for details).}
\label{fig:parameters}
\end{figure}

In Figure~\ref{fig:parameters}, the derived line widths, $\sigma_{\rm v}$, excitation temperatures, $T_{\rm ex}$, and line optical depths, $\tau$, are compared, with the one-to-one ratio shown by the dash-dotted line. {\dij Determining the optical depths and line widths are fairly straightforward and thus suffer from only a moderate uncertainty. The derived excitation temperatures, however, depend strongly on the calibration of the instrument and an assumption about the beam-filling nature of the emission. Thus the excitation temperatures should be taken as representative values with large, at least twenty percent, uncertainties.}
The top left panel shows a strong correlation in the measured line widths of the two nitrogen-bearing species, as discussed earlier in \S\,\ref{sec:obscomp:width}. The top right panel shows that a similar correlation can be found for the derived excitation temperatures of the two molecular transitions, although there is a hint of an offset to slightly higher, $\Delta T \sim 1$\,K, \nhhh\ temperatures. This result is intriguing given the large difference in critical densities of the two species observed. At densities sufficiently large to excite the \nnhp\ (1--0) transition, $n_c({\rm N}_2{\rm H}^+) \sim10^5\,$cm$^{-3}$, the \nhhh\ (1,1) transition should already be thermalized. Unless the gas density is much larger than this critical value, however, the \nnhp\ should be only sub-thermally excited. That the measured excitation temperatures are similar argues that the observed gas in which the \nnhp\ and \nhhh\ emission is arising may be extremely dense, $n \gg n_c({\rm N}_2{\rm H}^+)$. 

The bottom left panel in Figure~\ref{fig:parameters} plots the total line optical depths, {\dij integrated over all hyperfine components,} for the transitions observed. Although there is much scatter, the underlying trend is obvious, higher optical depths for \nhhh\ correlate with higher optical depths for \nnhp. More importantly, the total optical depths are fairly low - arguing that the peak optical depths in any particular hyperfine component are never much larger than unity
{\dij (for \nhhh\ the strongest hyperfine component has a weight of 0.5 while for \nnhp\  the strongest component has a weight of 0.26)}. Thus, the integrated line intensities in \S\,\ref{sec:abund:line} should not suffer significantly from optical depth effects. This result has been confirmed by comparing the integrated intensity in a single, isolated, hyperfine component against the total line intensity.

Given the expectation that the submillimeter continuum emission at 850\mum\ is directly related to the underlying column density of dust (and by extrapolation the total column density of gas - see \S\,\ref{sec:abund} and Appendix~\ref{sec:app}), it is possible to estimate the mean density within each core. In this analysis, we take the 850\mum\ emission, smoothed to a 30\arcsec\ beam, calculate the expected column density of dust and gas, and then divide by twice the measured core radius as given in Table 6 of \citet{kir07} for those cores which have radius measures (53 sources). The derived mean density within the core is only approximate but may be compared against the derived excitation temperatures to see if there are any obvious trends that might be due to sub-thermalization. The bottom right panel in Figure~\ref{fig:parameters} shows that for most of the cores the mean  density is $\left< n \right> \sim 2 \times 10^5\,$cm$^{-3}$. 

The strong correlation in excitation temperature and integrated line intensity, together with the equivalence of the kinematic features,  {\dij suggests} that the emission from the two nitrogen-bearing molecules is being observed from within a coincident dense region inside each core, whether pre-stellar or proto-stellar. 
{\dij In this situation, the true excitation temperatures are expected to be similar to the derived kinetic temperatures.
The fact that our measured $T_{\rm ex}$ values are often significantly lower than the $T_k \sim 11\,$K estimated by \citet{ros08} is somewhat puzzling, and suggests that our $T_{\rm ex}$ values may have been underestimated. This could result from beam dilution effects, if the sources are significantly smaller than the telescope beam, or from additional effects not considered in our analysis, like calibration problems or variations of the temperature along the line of sight.
Detailed modeling of the spatial distribution of $T_{\rm ex}$ and $T_k$ in a selected sample of cores is needed to
clarify this issue.}

\section{Chemical Properties of the Perseus Cores}
\label{sec:abund}

In the preceding sections, we have shown that there is a strong correlation between the observed properties of the two nitrogen-bearing species, \nnhp\ and \nhhh. The 850\mum\ submillimeter flux is also well understood as arising from dust emitting at a temperature $T_d \sim 11\,$K \citep{ros08}. In this section we use standard formulae for the conversion from observed continuum emission to \hh\ column density and from observed line parameters to \nnhp\ and \nhhh\ column densities in order to compare the abundances of these species source by source. Although the formulae used here to compute column densities have been presented in the literature before, we reproduce them in Appendix~\ref{sec:app} for completeness and to highlight some misconceptions that often creep into such calculations.

\subsection{Comparison of Line Strengths}
\label{sec:abund:line}

\begin{figure} 
\epsscale{0.8}
\includegraphics[width=1.0\textwidth]{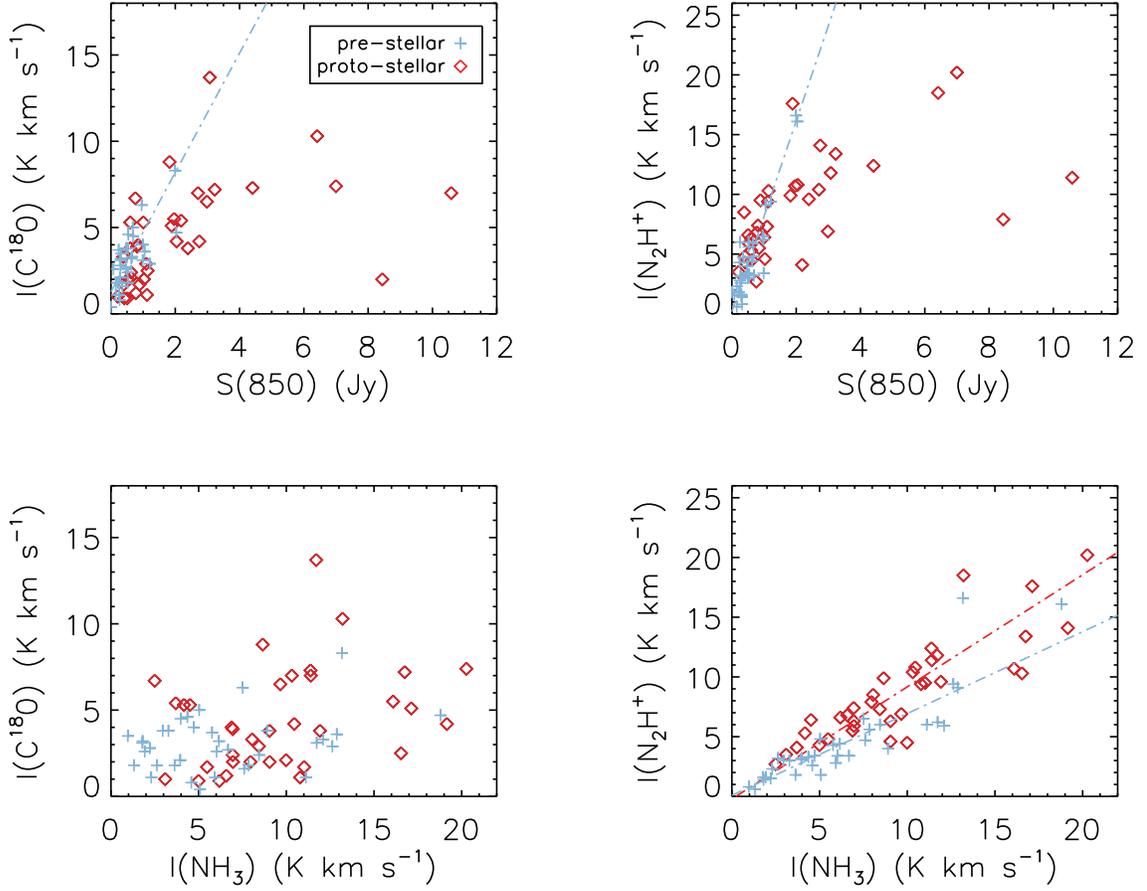}
\caption{Observed correlations in line intensity, I, and continuum strength, S, for the observed cores in Perseus. The symbols have the same meaning as in Figure \ref{fig:parameters}. Where useful, a best-fit linear relation is overlaid as a dash-dotted line with the line color denoting the underlying species being fit.}
\label{fig:intensity}
\end{figure}

Given the large number of cores in this sample, 71, it is useful to search for correlations in the measured intensities of the molecular lines, as well as against the strength of the continuum. Figure~\ref{fig:intensity} plots four of these correlations.  Where useful, a best-fit linear relation is overlaid as a dash-dotted line with the line color denoting the underlying species being fit. 

The top left panel in Figure~\ref{fig:intensity} shows the relationship between the 850\mum\ SCUBA flux  (smoothed to a 30\arcsec\ beam) and the \ceo\ (2--1) emission. There is a possible correlation for the pre-stellar cores, although the linear trend is driven by the few \ceo\ bright outliers. For the proto-stellar cores there is no obvious correlation. The top right panel shows the relation between the 850\mum\ SCUBA flux and the \nnhp\ (1--0) emission. Again, there is a possible correlation for the pre-stellar cores but no obvious trend for the proto-stellar cores. 
The bottom left panel in Figure~\ref{fig:intensity} reveals a lack of any correlation between the \ceo\ (2--1) and the \nhhh\ (1,1) emission, as expected from the discussion in \S~\ref{sec:obscomp:cent} and \ref{sec:obscomp:width}.  The bottom right panel, however, shows a very strong correlation between the \nhhh\ (1,1) intensity and the \nnhp\ (1--0) intensity, {\it for both pre-stellar and proto-stellar cores}.

\begin{figure} 
\epsscale{0.8}
\includegraphics[width=1.0\textwidth]{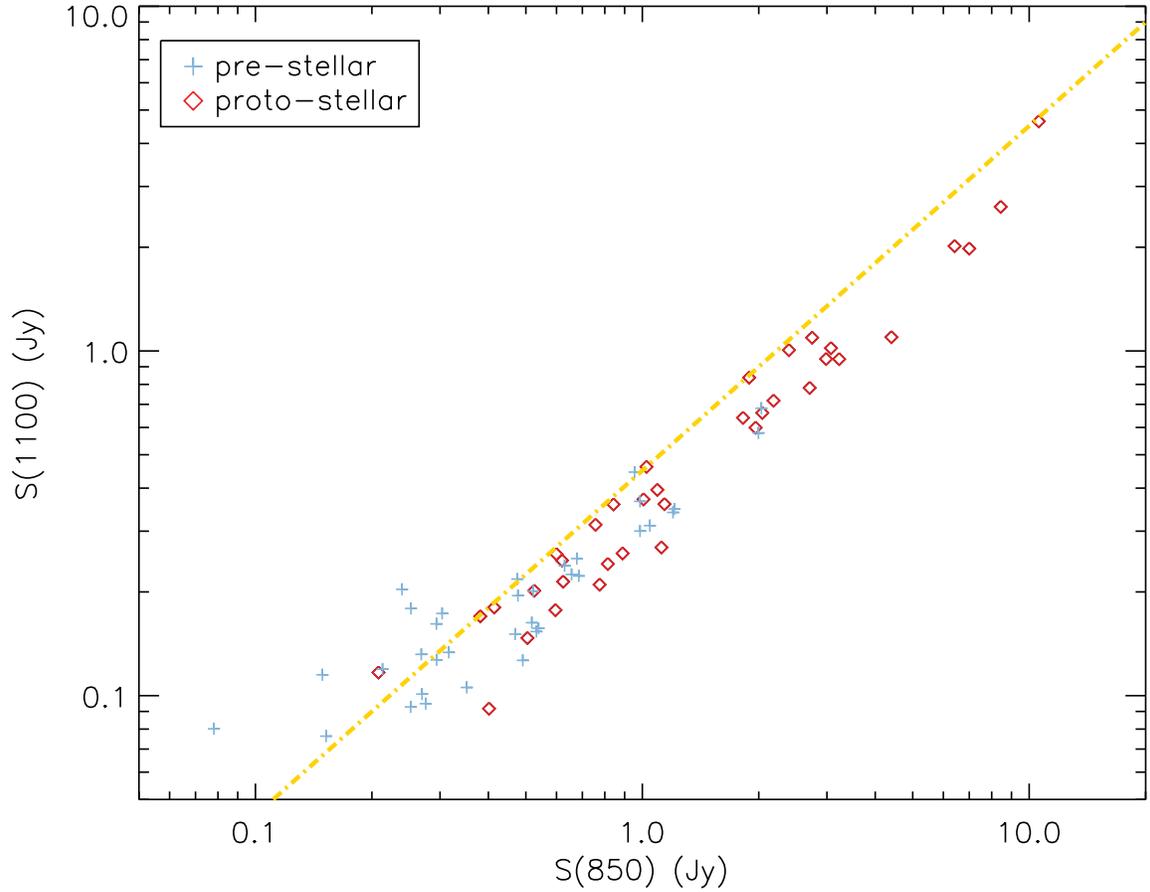}
\caption{Comparison of the submillimeter continuum flux within a 30 arcsecond beam at 850\mum\ and 1.1\,mm for the observed cores in Perseus. The predicted correlation, assuming a dust temperature $T_d = 11\,$K and a (sub)millimeter emissivity power-law $\beta =2$ is shown by the yellow line (see text for details). The symbols have the same meaning as in Figure \ref{fig:parameters}.}
\label{fig:continuum}
\end{figure}

For completeness, Figure~\ref{fig:continuum} shows the relationship between the 850\mum\ SCUBA flux and the 1.1\,mm Bolocam flux, both averaged over a 30\arcsec\ beam at the location of the \nhhh\ measurement. Note that this figure is presented using a logarithmic scale and that the two (sub)millimeter measurements correlate exceedingly well.  The straight (yellow) line through the data is {\it not} a fit but rather the expected correlation for dust at $T_d = 11\,$K, and with a (sub)millimeter emissivity power-law $\beta = 2$ \citep[for a discussion on $\beta$ see][and references therein]{joh06}. {\dij The typical kinetic temperature of the gas in these sources is $T_k \sim 11$K \citep{ros08}, and thus the agreement of the (sub)millimeter measurements with the 11\,K model imply that the gas and the dust are near thermal equilibrium, as expected for densities greater than $10^4\,$cm$^{-3}$ \citep{gol01}.}
A few of the bright proto-stellar sources have significantly ($\sim 50\,$\%) higher 850\mum\ fluxes than expected which may indicate {\dij moderate} warming toward these locations.  In \S\,\ref{sec:abund:column} where we determine column densities and abundances, the dust continuum measurements should be reliable to better than a factor of 2.

From the integrated intensity plots in Figure~\ref{fig:intensity}, it is evident that the \ceo\ (1--0) emission is poorly correlated with the dust emission observed in the (sub)millimeter (especially for proto-stellar cores) and uncorrelated with the nitrogen-bearing molecules.  For pre-stellar cores, there is a possible correlation of the nitrogen-bearing molecules and the dust. 

\subsection{Determination of Column Densities}
\label{sec:abund:column}

In general, the (sub)millimeter continuum emission observed from structure within molecular clouds is produced by radiating dust grains, and the emission is optically thin. Thus, if the dust temperature $T_d$ and the dust emissivity properties $\kappa_\lambda$ are known, the column density of dust is directly computable from the measured (sub)millimeter continuum flux. Usually, the dust is assumed to be well coupled to the gas, and the emissivity is given per unit gas and dust. The required equation is presented in Appendix~\ref{sec:app:dust}. The derived column density of \hh, {\dij $N$(\hh)}, toward each pointing is provided in Table~\ref{tab3}, assuming that the dust temperature is $T_d = 11\,$K, the typical ammonia temperature measured by \citet{ros08}. Raising the dust temperature to $T_d = 16\,$K halves the measured column density, while a dust temperature of $T_d = 26\,$K is needed to quarter the measured column density values. 

Assuming the molecular emission from a given transition is optically
thin, the integrated line strength should be proportional to the
column density of the emitting gas (see Appendix~\ref{sec:app:mol}).
If the gas is moderately optically thick, and the optical depth can be
estimated, correction factors can be utilized to reconstruct the total
column density. These calculations, however, are extremely dependent
on the state of the gas, including the kinetic temperature, the
density of collision partners, and the equilibrium properties of the
molecule. For the two nitrogen-bearing species, \nnhp\ and \nhhh, the
hyperfine structure of the emission provides a reasonable measure of
the optical depth, which coupled with the peak intensity of the line
yields an estimate of the excitation temperature of the transition (in
practice a more sophisticated fit to the hyperfine structure is
utilized). Thus, the number of molecules radiating in this transition,
along the line of sight, can be deduced. In order to determine the
total abundance of the molecule, however, the energy partition
function must be constructed and the level abundances computed. 
Additionally, for some molecules, including \nhhh, the energy levels
are divided between ortho and para forms, which can be treated as
independent species because they are not connected by normal radiative
or collisional transitions. As our observations concern only 
\nhhh(1,1) and (2,2), in this paper we will only consider the para form
of ammonia (p-\nhhh). 

In Table~\ref{tab3}, we present the derived column densities for \nnhp\ and p-\nhhh, utilizing the formulae presented in Appendix~\ref{sec:app:mol}. The adopted physical properties, $\sigma_{\rm v}$, $T_{\rm ex}$, and $\tau$ are taken from the physical parameters fit to each spectrum (see Table~\ref{tab2} and \S~\ref{sec:obscomp:tex}).  Additionally, determination of the conversion factor, $N/N_l$, from the column density of the observed state, $N_l$, to total column density, $N$, (see Appendix~\ref{sec:app:mol2}) requires an assumption that the energy levels within each molecule are in equipartition at an adopted excitation temperature. As discussed in Appendix~\ref{sec:app:mol2}, the conversion factor necessary for \nhhh\ (1,1) is relatively independent of the adopted excitation temperature, due to the large energy gap between the (1,1) and (2,2) states. We assume the conversion factor for \nhhh\ is 2, which is valid to within 10\% for temperatures less than $15\,$K. The conversion factor for \nnhp, however, is quite dependent on the assumed excitation temperature used in the partition function.  In the column density analysis, the derived excitation temperatures for the \nnhp\ (1-0) transition were averaged over all pre-stellar and proto-stellar cores to derive effective values of $T_{\rm ex} = 5.7\,$K and $T_{\rm ex} = 6.9\,$K, respectively. Thus, the conversion to total column density for \nnhp\ for pre-stellar cores is $2.9$, and for proto-stellar cores the conversion is $3.4$.

A test for the validity of the column density ratio measurements between \nhhh\ and \nnhp\ was performed using a Monte Carlo radiative transfer code based on that of \citet{ber79} and as discussed in more detail by
\citet{taf04}.  We produced model cores with realistic physical conditions for the density distribution, central dust column densities similar to those observed for the Perseus core sample, and internal temperatures of 11\,K. Synthetic spectra were calculated for various input abundances of  \nhhh\ and \nnhp, and these spectra were reduced using the same procedures as for the Perseus observations. 
The derived column densities and abundances were found to accurately reflect the input values to within about 10\%.

In \S~\ref{sec:obscomp:tex}, the excitation temperatures for both nitrogen-bearing species were seen to be similar, suggesting that the density in the emitting region might be larger than the critical densities of both molecules. Under this assumption, the true excitation temperature for the observed transitions would be $\sim T_k (\sim 11\ {\rm K})$, as both lines should be thermalized. Deviations in the observed $T_{\rm ex}$ from $T_k$ would then be due to beam dilution effects.  Under this scenario, the conversion to total column density of \nnhp\ would be $N/N_0 = 5.3$, or 83\% larger than the value we have adopted for pre-stellar cores and 56\% larger for proto-stellar cores. In both cases the uncertainty introduced is less than a factor of 2. The conversion to total column density of \nhhh\ would remain 2, however.

\subsection{\nnhp\ and p-\nhhh\ Relative Abundance}
\label{sec:abund:n}

\begin{figure} 
\epsscale{0.8}
\includegraphics[width=1.0\textwidth]{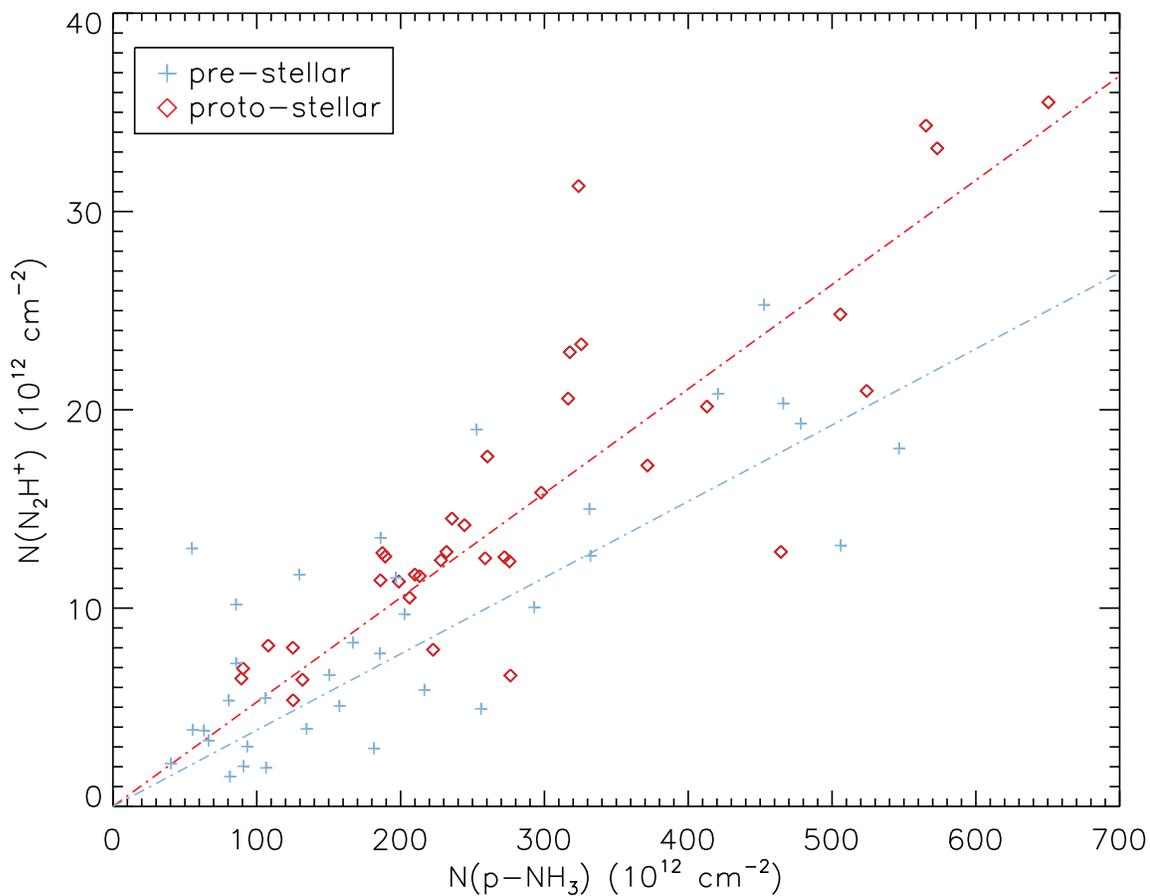}
\caption{Calculated column densities of p-\nhhh\ versus \nnhp\ for the observed cores in Perseus. Note that the plot has linear axes. The symbols have the same meaning as in Figure \ref{fig:parameters} and the best linear fits to the data are overlaid as dash-dotted lines with the line color denoting the underlying species being fit. The best fit slopes are $N$(p-\nhhh)/$N$(\nnhp) $=25 \pm 12$ and $20 \pm 7$ for the pre-stellar and proto-stellar cores, respectively.}
\label{fig:column_den}
\end{figure}

Figure~\ref{fig:column_den} shows the strong correlation between the column density of \nnhp\ and p-\nhhh. The tight fit for the two species can be anticipated from the correlation for line intensities seen in Figure~\ref{fig:intensity} (see also Appendix~\ref{sec:app:mol}), and the fact that the emission lines are never particularly optically thick. The scaling from observed line intensity to column density depends both on observing methods (telescope efficiencies and corrections for atmospheric attenuation) as well as corrections for the measured physical conditions within the gas (e.g., excitation temperatures).  Each of these measures has uncertainty associated with it and should dilute any observed underlying column density correlations, as the \nnhp\ (1--0) and \nhhh\ (1,1) transitions were measured at different times and with different telescopes. As well, two different fitting programs were used to determine the physical parameters from the hyperfine components. It is more likely that there are correlated uncertainties in the integrated intensity of a given species (e.g., an error in the assumed telescope efficiency) and thus the scaling for {\it all} the column densities of either \nnhp\ or p-\nhhh\ may be off by about 20\%.

The top panels of Figure~\ref{fig:abundance} plot the abundance of p-\nhhh\ with respect to \nnhp\ as a function of the total \hh\ column density (derived from the dust continuum measurements) and as a function of the para-ammonia column density. As in all other plots, the blue plus signs represent pre-stellar cores and the red diamonds represent proto-stellar cores. The scatter in the abundance ratio of these two nitrogen-bearing species is extremely small - less than a factor of two about the mean value. As well, there is only a small hint at a difference in this ratio for the pre-stellar and proto-stellar cores. For the complete sample of 71 cores for which measurements can be made, an average abundance ratio of $22 \pm 10$ is found for p-\nhhh\ versus \nnhp.  Separating the 35 pre-stellar and 36 proto-stellar cores yields only small variations in this abundance ratio, $25 \pm 12$ and $20 \pm 7$ respectively (see also Figure~\ref{fig:column_den}).

Given that \nnhp\ is a highly reactive molecular ion, and is known to form readily in regions where CO depletes \citep[see for example,][]{ber07}, it would appear that the formation of p-\nhhh\ must follow a similar route to the formation of \nnhp.  No discernible p-\nhhh\ was detected from the moderate density bulk gas surrounding the cores, otherwise, an observational correlation with the line widths of the \ceo (2--1) transition should have been noted. As well, within the core the line strengths for both nitrogen-bearing species must be dominated by dense gas emission in order for the excitation temperatures to be similar (see \S~\ref{sec:obscomp:tex}). Thus, it would appear that both \nnhp\ and p-\nhhh\ are being produced in the densest regions of the cores, in zones where the CO is most likely to be freezing out.

Perhaps more interestingly, \nnhp\ is known to be destroyed by CO and thus in proto-stellar cores one might expect that the abundance of \nnhp\ should decrease as the internal warming evaporates frozen CO back into the gas. No such destruction of p-\nhhh\ is postulated, however, and thus the lack of any significant change in the relative abundance of these two species between pre-stellar and proto-stellar cores suggests that the warming zone is limited to a small volume within the core. We return to this discussion on chemistry in \S\,\ref{sec:disc}.

\subsection{\nhhh\ and \hh\ Relative Abundance}
\label{sec:abund:h}

The bottom panels of Figure~\ref{fig:abundance} plot the abundance  of p-\nhhh\ with respect to \hh\ as a function of the total \hh\ column density and as a function of the para-ammonia column density.  For the pre-stellar cores there is only a hint that the abundance ratio is not constant at $\sim 10^{-8}$. The proto-stellar cores, however, show a clear trend of lower ammonia abundance in higher column density cores. The trend is reminiscent of the integrated intensity plot shown in the top left panel of Figure~\ref{fig:intensity}, where at high 850\mum\ flux levels (i.e. high \hh\ column densities) the integrated intensity in the \nhhh\ (1,1) line appears to saturate. As noted in \S\,\ref{sec:obscomp:tex}, this is not an optical depth effect; the hyperfine structure of the \nhhh\ does not show evidence of reaching the necessarily high optical depths. {\dij A similar effect has also been noted for \nnhp\ versus \hh\ for cores in Ophiuchus by \citet{fri09}.}

We remind the reader that the \hh\ column density is derived from dust continuum measurements. Consideration of Equation~\ref{e:nh2} suggests that significant heating of the proto-stellar cores would produce enhanced emission which might be misinterpreted as higher column density if the appropriate higher temperature is not used in the analysis. Low mass protostars, however, do not have sufficient luminosity to heat large portions of their envelope {\dij \citep[see for example,][]{jor06b, sta07}}. As well, there is no evidence for significant warming of the envelope in the kinetic temperature determinations derived from the \nhhh\ (1,1) and (2,2) lines \citep{fos09}. Finally, we note that for the brightest 850\mum\ sources, the calculated under-abundance of p-\nhhh\ is greater than an order of magnitude, requiring an extreme change in temperature, or other dust properties. We return to this issue in \S\,\ref{sec:disc}.

\section{Discussion of the Results}
\label{sec:disc}

From the observations of \nnhp, \nhhh, \ceo, and dust continuum emission, we have uncovered a tight correlation between the abundance of the two nitrogen-bearing species for both pre-stellar and proto-stellar cores in the Perseus molecular cloud. No similar correlation is found for the nitrogen-bearing species versus \co. As well,  only the pre-stellar cores show evidence for a constant abundance of these species versus \hh. In this section, we consider the relevant time scales for chemical and core evolution, in order to better understand the possible processes by which the correlations are produced. A more detailed discussion of these processes can be found in the review paper by \citet{ber07}.

\subsection{Chemical Evolution}
\label{sec:disc:time}

While the entire chemical pathways to the formation of \nnhp\ and
\nhhh\ within molecular clouds are not fully understood, there is
general agreement on the expected steps. For \nnhp, destruction via
interactions with \co\ keeps the abundance of this species low until
the \co\ freezes out onto dust grains \citep{cas99, cas02, taf02}. Once the \co\ has
frozen out substantially, the abundance of \nnhp\ can increase
significantly in the gas phase through interactions between N$_2$ and
H$_3^+$. At the low temperatures expected deep within molecular
clouds, \co\ efficiently freezes to dust and thus the relevant
time scale for \co\ depletion is the collision time with dust:
\begin{equation}
t_{\rm{co,dep}} \sim 10^4 \left({ 10^5\,{\rm  cm}^{-3} \over {n_{\rm H_2}}}\right) \ {\rm yr}.
\end{equation}

For \nhhh\ there is no obvious destruction mechanism with \co\ and thus it is
possible to have both simultaneously, assuming that there is sufficient time to
produce the parent product \nn\ in the bulk cloud.  Thus, the absence of \nhhh\ 
in the extended (\ceo\ emitting) cloud gas suggests that the \nhhh\ formation 
time scale is longer than the CO formation time scale, and comparable to that 
of core formation and CO freeze out. 
Indeed, \nhhh\ (and \nnhp) are considered ``late time"  molecules 
due to the long time needed to activate their nitrogen chemistry, which starts with 
a slow neutral-neutral reaction \citep[e.g.][]{suz92}. Despite this similar starting point
in their production, the different behavior of \nhhh\ and \nnhp\ with respect to the
presence of CO in the gas phase makes somewhat surprising the strong correlation 
found between the abundance of these species over the full range of core properties 
seen in Perseus. It should be noted that a favored \nhhh\ formation mechanism starting 
with the electronic recombination of \nnhp\ \citep{gep04, aik05} has been proven
inefficient by \citet{mol07}, so a new generation of chemical models taking these
recent measurements into account is clearly needed.

\subsection{Pre-Stellar Core Evolution}
\label{sec:disc:pre}

The formation time for a pre-stellar core depends on the physical
processes responsible for bringing the material together and is still
strongly debated in the community. Before the onset of collapse,
however, the fastest rate at which the core can be assembled may be
estimated from the observed physical parameters of the material around
the core. The maximum infall rate onto the core is
\begin{equation}
\dot M_{\rm max} \sim 4\pi R^2 v\left< n \right>\mu m_{H},
\end{equation}
where $m_{H}$ is the mass of a hydrogen atom, $\mu$ is the mean
molecular weight, and, for the cores in Perseus, $\left< n \right>
\sim 5 \times 10^3\,$cm$^{-3}$ is the mean density in the region {\it
surrounding} the core, $R \sim 5 \times 10^{16}\,$cm is the typical
core size, and $v \sim 0.4\,$km\,s$^{-1}$ is the typical non-thermal
line width observed in \ceo\ (in general one does not expect that
this velocity gradient is due entirely to infall but using it yields a
reasonable upper limit for the infall rate). Thus, even if the
pre-stellar core is assembled at the maximum rate,  a typical Perseus
0.5\,$M_\odot$  core takes at least
\begin{equation}
t_{\rm form} \ga 1.4 \times 10^5\ 
\left( { R \over 5 \times 10^{16}\ {\rm cm}}\right)^{-2}
\left( { v \over 0.4\ {\rm km}\,{\rm s}^{-1}} \right)^{-1}
\left( { n \over 5 \times 10^3\ {\rm cm}^{-3}} \right)^{-1}\
{\rm yr}.
\end{equation}
This time scale is significantly longer than the depletion time for
\co\ within the core, where the density is more than an order of
magnitude higher. Thus, it would appear that the chemical enhancement
of \nnhp\ should proceed in lock-step with the increase in core mass
during the pre-stellar core phase as observed (see bottom left panel in
Figure~\ref{fig:abundance}).

\subsection{Proto-Stellar Core Evolution}
\label{sec:disc:proto}

Once the core becomes sufficiently massive, gravity will dominate over
internal thermal pressure and the core should collapse on a free-fall
time
\begin{equation}
t_{\rm coll} \sim 10^5 \left({ 10^5\,{\rm  cm}^{-3} \over {n_{\rm H_2}}}\right)^{1/2} \ {\rm yr}.
\end{equation}
It is important to note that the time for collapse is longer than the
freeze-out time at the typical densities inside cores, $n_{\rm H_2}
\sim 10^5\,$cm$^{-3}$, and that the time scale for freeze-out drops
faster than the collapse time as the core density increases. Thus, even if
the core were to begin collapse before \co\ depletion and chemical
enrichment of the \nnhp\ and \nhhh, these nitrogen-bearing species
should be substantially enhanced before the collapse becomes advanced.

\subsection{The Dichotomy for Dense Gas Tracers in Proto-Stellar Cores}
\label{sec:disc:dich}

\begin{figure} 
\epsscale{0.8}
\includegraphics[width=1.0\textwidth]{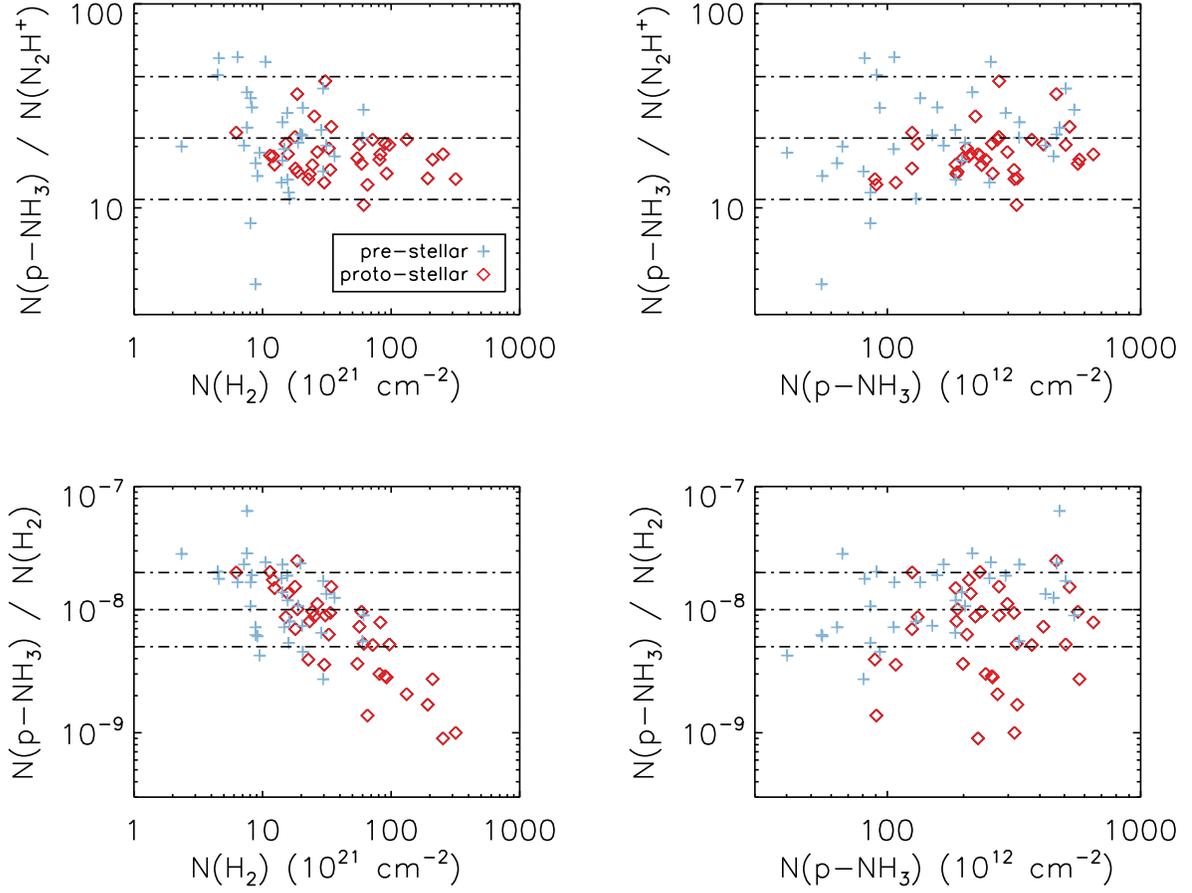}
\caption{Comparison of the abundance ratios between  \nnhp, p-\nhhh, and \hh\ for the observed cores in Perseus. The symbols have the same meaning as in Figure \ref{fig:parameters}. The top panels show the nitrogen-bearing species abundance ratio against the total column of \hh\ and p-\nhhh. Here, the central horizontal line marks the mean value of the abundance ratio, 22,  while the additional horizontal lines indicate a change in the ratio by a factor of 2, up or down. The lower panels show the p-\nhhh\ to \hh\ abundance ratio against the total column of  \hh\ and p-\nhhh. Here, the central horizontal line marks the rough value of the abundance ratio, $10^{-8}$,  while the additional horizontal lines indicate a change in the ratio by a factor of 2, up or down. Note that the only obvious non-constant ratio is found for the $N$(p-\nhhh)/$N$(\hh) in proto-stellar cores (red diamonds) versus $N$(\hh) (bottom left panel).}
\label{fig:abundance}
\end{figure}

While the three dense gas tracers, \nnhp, \nhhh, and \hh\ [traced by the (sub)millimeter dust continuum], observed in Perseus appear to mimic one another for pre-stellar cores, the trend found between the nitrogen-bearing species and \hh\ in Figure~\ref{fig:abundance} for proto-stellar cores requires a significant drop in the abundance of \nhhh\ and \nnhp\ with increasing \hh\ column. {\dij A similar trend might exist in the pre-stellar core sample but without high column density sources this cannot be confirmed}. Had the results been reversed, with the pre-stellar cores showing a {\dij large} range of abundance ratios then one might have appealed to a chemical differentiation during core formation. However, assuming that all proto-stellar cores began their lives similarly to the observed homogeneous pre-stellar cores, gravitational collapse alone (\S\,\ref{sec:disc:proto}) does not appear to provide a mechanism for increasing the mass of \hh\ without also increasing the mass of the nitrogen-bearing species. A number of processes may be invoked to solve this dichotomy and these are described below. Unfortunately, it is not possible to definitively choose which process is most likely, as all have strengths and weaknesses.

One possible solution might invoke the destruction of nitrogen-bearing molecules.  In the inner region around the proto-stellar core the gas temperature will be raised due to the heating from the protostar. Where the gas temperature exceeds $\sim 35-40\,$K the \co\ is observed to evaporate from the grains
\citep{jor05a, jor02, jor04, dot04} and re-enter the gas phase where it will destroy the \nnhp. The size of this zone, however, is much smaller than the core and thus this effect should be marginal within the single-dish observations presented here. For low-mass protostars, the inner warm zone is only $\sim 100\,$AU compared with the $\ga 5000\,$AU core radii, and $\sim 7500\,$AU beam diameters. If the dust evaporation temperature is significantly lower, the evaporation envelope may grow much larger. In that case, however, the warmed gas will be rich in \co, providing a strong correlation between dust continuum and \co\ emission for the brightest sources, which is not observed in Perseus. Additionally, while the \nnhp\ abundance will decrease in the inner warm region, there is no clear reason why the \nhhh\ abundance should also decrease.

A second possible mechanism for changing the relative abundances of the nitrogen-bearing species versus \hh\ in proto-stellar cores is for the heavier molecules to freeze-out onto the dust particles. Chemical models predict that these species should eventually freeze-out, however, there has only been a little observational evidence of this effect, especially for \nnhp \citep[see][and references therein]{ber02,pag07}.  In order for this scenario to explain the observations in Perseus, the collapsing proto-stellar cores would need to be accumulating new material in their outer envelopes, where \co\ freezes out and \nnhp\ and \nhhh\ form, while in the interior even these heavy species would be depleting onto dust grains.  In this manner, the column of \hh\ would increase over time while the total column of the nitrogen-bearing species would remain bounded.

A third, entirely different explanation for the proto-stellar trend seen in Figure~\ref{fig:abundance} is also plausible. While the conversion from emission line strength to total molecular column density is relatively straightforward, and apparently validated by the excellent agreement between the \nnhp\ and \nhhh\ abundances, the conversion from (sub)millimeter dust continuum emission to \hh\ column density requires knowledge of both the dust temperature and the dust emissivity. As mentioned earlier, neither of these values is expected to vary dramatically within cores - typical uncertainties in the conversion from observed emission to column density for each is a factor of 2, much smaller than the factor of 10 variation seen in the observations.  We note, however, that it is possible that these uncertainties are correlated and that, as the central temperature increases in the core, the dust properties also change. Such a scenario could  significantly decrease the variation in abundance ratio between the nitrogen-bearing species and \hh\ observed for the Perseus cores. It would, as well, fundamentally affect the manner in which proto-stellar envelope masses are determined. 

There is a clear need for the observers and modelers to work together to solve this dichotomy between the dense gas tracers if we are to achieve a self-consistent description of the proto-stellar environment.

\section{Conclusions}
\label{sec:concs}

We have investigated the observed chemical properties of 35 pre-stellar and 36 proto-stellar cores in 
the nearby Perseus  molecular cloud. By combining spectroscopic observations of \nnhp\ (1-0), \ceo\ 
(2-1), and \nhhh\ (1,1) along with (sub)millimeter continuum flux measurements, we are able to determine correlations in the properties of the emitting regions for each species, and the abundance ratios between the various chemical tracers. The main conclusions from our investigation are:

{\noindent\bf(1) }The kinematic properties of \nhhh\ and \nnhp\ are extremely similar and quite different from the kinematic properties of the \ceo\ molecule, strongly suggesting that the formation and destruction of these two nitrogen-bearing species are well coupled.

{\noindent\bf(2) }For all cores the abundance ratio between the two nitrogen-bearing species is fixed at $N$(p-\nhhh)/$N$(\nnhp) = $22 \pm 10$. Dividing the cores into pre-stellar and proto-stellar samples does not result in significantly different abundance ratios, reinforcing the notion that the two nitrogen-bearing species trace the same gas and chemically evolve together.

{\noindent\bf(3)} For pre-stellar cores the abundance ratio between p-\nhhh\ and \hh\ is fixed at $N$(p-\nhhh)/$N$(\hh) $\sim 10^{-8}$, where the \hh\ column density is derived from the (sub)millimeter dust continuum measurements. This reinforces the notion that observations of \nhhh, \nnhp, and (sub)millimeter emission all trace the same dense gas and may be used interchangeably when searching for and analyzing pre-stellar cores in molecular clouds.

{\noindent\bf(4)} For proto-stellar cores the abundance ratio between the nitrogen-bearing species and \hh\ declines from the pre-stellar value as the column of \hh\ increases, where again the \hh\ column density is derived from the (sub)millimeter dust continuum measurements. This result suggests that observers should be careful when using a single dense gas tracer (\nhhh, \nnhp, or (sub)millimeter emission) to determine the properties of proto-stellar cores. The monotonic trend in the observed abundance ratio with \hh\ column density may indicate a simple underlying physical explanation, although decoupling the various possibilities outlined in \S\,\ref{sec:disc:dich} is likely to be tricky and requires further careful observations and detailed modeling.

\acknowledgements
\section{Acknowledgments}
\label{sec:ack}

We thank James Di Francesco, Jon Swift, Paola Caselli, and the anonymous referee for helpful discussions.

Doug Johnstone and Erik Rosolowsky are supported by Natural Sciences and Engineering Research Council of Canada (NSERC) Discovery Grants. Helen Kirk is supported by a University of Victoria Fellowship.

We acknowledge the use of data from the following observational facilities: CSO, GBT, IRAM, JCMT, Palomar, {\it Spitzer\ }, and 2MASS.

\newpage

\begin{appendix}
\section{Appendix - Formulae For Determining Column Densities and Abundances}
\label{sec:app}

In this section, we first discuss the underlying relationship between
the observed submillimeter continuum brightness of the cores and the
column density of \hh. Next we show how the integrated line
intensities, or physical properties of \nnhp\ and \nhhh\ can be used
to measure column densities.

\subsection{Total Gas Column Density from Submillimeter Continuum Emission}
\label{sec:app:dust}

Assuming a constant dust temperature $T_d$ and optically thin
conditions, the column density of gas, $N_{\mathrm{H}_2}$, can be related to
the 850\mum\ submillimeter continuum brightness within a beam,
$S_{850}$, by
\begin{equation}\label{e_submm_bright}
N_{H_2} = S_{850} \left[ \Omega_{\rm bm}\, \mu\, m_{\mathrm{H}}\, \kappa_{850}\,
  B_{850}(T_d) \right]^{-1}.
\end{equation}
In the above equation, $\Omega_{\rm bm}$ is the solid angle subtended
by the observation, $m_{\mathrm{H}}$ is the mass of atomic hydrogen, $\mu = 2.37$
is the mean molecular weight, $\kappa_{850}$ is the dust opacity per
unit mass column density (gas plus dust) at $850\, \mu$m, and
$B_{850}$ is the Planck function evaluated at $850\,
\mu$m. Substituting typical values for these quantities, and assuming
a 30 arcsecond beam, yields 
\begin{equation}
\label{e:nh2}
N_{H_2} = 3.0 \times 10^{22} 
		\left[ S_{850} \over {\rm 1\ Jy\ (30'' \,bm)}^{-1} \right]
		\left[ \kappa_{850} \over 0.02\ {\rm cm}^{-2}\ {\rm g}^{-1} \right]^{-1}
		{\left[ \exp\left({17\, {\rm K} \over T_d} \right) - 1 \right] \over
		 \left[ \exp\left({17\, {\rm K} \over 11\,{\rm K}} \right) - 1 \right]}
		{\rm cm}^{-2}.
\end{equation}

Note that the measured column density scales linearly with the
submillimeter continuum brightness, $S_{850}$, and inversely with the
dust opacity, $\kappa_{850}$. While the appropriate value for
$\kappa_{850}$ is still uncertain \citep[see for example][]{van99}, the ranges of values
used in the literature span only about a factor of 2. In this paper we take $\kappa_{850} = 0.02\
{\rm cm}^2$ as a fiducial value.

Only the dust temperature $T_d$ enters the equation in a non-linear manner. For the
sources observed in Perseus, however, the \nhhh\ kinetic temperature
measurements suggest that the dense gas temperature (which should
couple extremely well with the coexistent dust temperature) is
 narrowly scattered around 11~K.

\subsection{Molecular Species Column Density}
\label{sec:app:mol}

To determine the total column density of a molecule, we first need
to calculate the column density in the observed transition, $N_{l}$
(where the subscript $l$ refers to the lower energy state of the
observed transition), and then, through use of an appropriate
partition function, calculate the total column density of the species,
$N$.

\subsubsection{Column Density in the Observed Transition}

The column density in the lower energy state $l$ is determined by
assuming statistical equilibrium and expressing the optical depth in
terms of the column density in the lower state \citep[e.g.,][]{roh04}.
This relationship is then inverted yielding an expression for $N_l$:\begin{equation}
\label{e:Nm}
N_{l} = { 8\pi\nu_{ul}^2 \over c^2 }\, { g_{l} \over g_{u}}\, { 1 \over A_{ul}} 
		\left[ 1 - \exp \left( -{h\,\nu_{ul} \over k\,T_{{\rm
                        ex}}} \right) \right]^{-1}\int \tau(\nu)\,d\nu,
\end{equation}
where $c$ is the speed of light, $h$ is the Planck constant, $k$ is
the Boltzmann constant, $\nu_{ul}$ is the frequency of the ($u,l$)
transition, $g_{l}$ and $g_{u}$ are the statistical weights, $A_{ul}$
is the Einstein $A$ coefficient, $T_{{\rm ex}}$ is the excitation
temperature of the transition, and the frequency-integrated line
optical depth is $\int \tau(\nu)\,d\nu$.

For both species in our analysis, p-NH$_3$ and \nnhp, the observed
transitions have resolved hyperfine structure.  Hence, $\tau(\nu)$
must reflect the integrated properties of the $i$ hyperfine components
via
\begin{equation}
\label{e:optdepth}
\tau( \nu) = \tau_{ul} \displaystyle\sum_{i=1}^n\,s_i
		\exp\left[ - {  \left( \nu - \nu_i - \nu_{\rm lsr}\right) ^2 \over 2 \, {\sigma_{\nu}}^2 } \right].
\end{equation}
where $\tau_{ul}$ is the total optical depth in the line as determined
by fits to the observed line profile, $s_i$ is the weight of the $i$th
hyperfine component defined such that $\sum s_i =1$, $\nu_i$ is the
rest frequency of the $i$th hyperfine transitions and $\nu_{LSR}$ is
the frequency shift induced by the systemic motions of the gas.  In
Equation \ref{e:optdepth}, the gas responsible for line formation has
been assumed to have a Gaussian distribution of line-of-sight motions.
The frequency width of each hyperfine component is then given by the
Doppler relationship:
\begin{equation}
\sigma_{\nu} = \frac{\sigma_{\mathrm{v}}}{c}{\nu_{ul}},
\end{equation}
and is, to high precision, the same for all the hyperfine transitions.
Hence, we characterize the line with a single rest frequency
$\nu_{ul}$.  Using Equation \ref{e:optdepth}, the integral in Equation
\ref{e:Nm} can be evaluated yielding
\begin{equation}
\int \tau(\nu)\, d\nu = (2\pi)^{1/2}\, \tau_{ul}\, \sigma_{\nu}.
\end{equation}

The relative strengths of the hyperfine components of the transitions
allow for a unique determination of the optical depth $\tau_{ul}$
unless the optical depth is very low \citep[see for
  example,][]{ros08}.  Furthermore, the line widths provide a reliable
measure of $\sigma_{{\rm v}}$ and the line intensities yield $T_{{\rm
    ex}}$ through the observed telescope main beam temperature $T_{\rm
  mb}$:
\begin{equation}\label{e:Tmb}
T_{{\rm mb}}(\nu) = \eta_f \left[ J(T_{{\rm ex}}) - J(T_{\rm bg}) \right]
					\left[1- e^{-\tau(\nu)}\right],
\end{equation}
where $\eta_f$ is the fraction of the telescope beam filled by
emission, $T_{\rm bg} = 2.73$~K and
\begin{equation}
J(T) = { T_{ul} \over \exp\left({T_{ul}/T}\right) - 1}.
\end{equation}
Here $T_{ul}\equiv (h\nu_{ul}/k)$, the equivalent temperature of the
transition energy.  $T_{{\rm mb}}$ is related to the observed
atmosphere corrected antenna temperature $T^*_A$ by $T^*_A = \eta_{\rm
  mb}(\nu_{ul})\, \eta_f T_{{\rm mb}}$ where $\eta_{\rm mb}
(\nu_{ul})$ is the main beam efficiency of the telescope at frequency
$\nu_{ul}$.

With these observed properties, Equation \ref{e:Nm} can be further
simplified to yield
\begin{equation}\label{e:Nm_2}
N_{l} = \left[ {4\,(2\,\pi)^{3/2} \over c^3 } \, { g_{l} \over g_{u}}\,  { {\nu_{ul}}^3 \over A_{ul}} \right]
            \left[ 1 - \exp \left( -{T_{ul} \over T_{{\rm ex}} } \right) \right]^{-1}\,
            \tau_{ul}\, \sigma_{{\rm v}}.
\end{equation}

It is often convenient to calculate the integrated intensity of the
line, $I_{ul}$, as this is a robust observational measure. Taking
Eqn \ref{e:Tmb}, converting from frequency to velocity space, and
integrating over all velocities yields
\begin{equation}
I_{ul} = \left[ J(T_{{\rm ex}}) - J(T_{\rm bg}) \right] \int
\left[1- e^{-\tau(\mathrm{v})}\right] d\mathrm{v}.
\end{equation}
At low optical depths this converts directly to
\begin{equation}\label{e:Im_thin}
I_{ul} =  (2\,\pi)^{1/2}\, \left[ J(T_{{\rm ex}}) - J(T_{\rm bg}) \right]\,  
           \tau_{ul}\, \sigma_{{\rm v}}, \quad \tau_{ul} \ll 1.
\end{equation}
The last term in the above equation is identical to the last term in Eqn \ref{e:Nm_2} suggesting that we combine the two equations
\begin{equation}
I_{ul} = \delta(\tau_{ul}, \sigma_{{\rm v}}) \,
		\left( { 8\,\pi \over c^3 } \, { g_{l} \over g_{u}}\,  { {\nu_{ul}}^3 \over A_{ul}} \right)^{-1}\,
		\left[ J(T_{{\rm ex}}) - J(T_{\rm bg}) \right]
		 \left[ 1 - \exp \left( -{T_{ul} \over T_{{\rm ex}} } \right) \right]\, N_{l}
\end{equation}
where $\delta(\tau_{ul}, \sigma_{{\rm v}})$ is a measure of the
deviation between the extrapolation of the optically thin line
intensity equation (Eqn \ref{e:Im_thin}) and the exact
calculation. Explicitly,
\begin{equation}
\delta(\tau_{ul}, \sigma_{{\rm v}}) = { \int  
			\left[ 1 - e^{-\tau(\mathrm{v})} \right]~d\mathrm{v}
			\over \int \tau(\mathrm{v})~d\mathrm{v}}.
\end{equation}
This formulation in terms of an escape probability term ($\delta$) is
useful since it allows the estimation of column density ratios between
two species in terms of their integrated intensities.  We
append subscripts $m$ and $n$ to indicate quantities for the two
species and their respective observed transitions:  
\begin{equation}
\label{e:iratio}
{I_{ul,m} \over I_{ul,n}}  =
		\left[{\delta_m \over \delta_n}\right]
		\left[{J_m(T_{{\rm ex},m}) - J_m(T_{\rm bg}) \over J_n(T_{{\rm ex},n}) - J_n(T_{\rm bg})}\right]
		\left[{1 - \exp(-T_{ul,m}/T_{{\rm ex},m}) \over 1 - \exp(-T_{ul,n}/T_{{\rm ex},n}) }\right]
		\left[  D_{ul,m} \over D_{ul,n} \right]
		\left[ {N_{ul,m} \over N_{ul,n}}Ê\right].
\end{equation}
where $D_{ul,m}$ is a fixed number, dependent only on the physical
properties of the molecular transition;
\begin{equation}
D_{ul,m} = 		{ g_{u,m} \over g_{l,m}}\,  { A_{ul,m} \over {\nu^3_{ul,m}}}.
\end{equation}
Provided that $T_{{\rm ex}}\gg T_{ul}$ for both species and $T_{{\rm
    ex},m}\approx T_{{\rm ex},n}$ then the second
bracketed term in Equation~\ref{e:iratio} is approximately unity and
the third bracketed term is reasonably approximated as
$T_{ul,m}/T_{ul,n}$, thus rearranging we get
\begin{equation}\label{e_ratio}
 {N_{l,m} \over N_{l,n}} =
		\left[{T_{ul,n}\,D_{ul,n} \over T_{ul,m}\,D_{ul,m}}\right]
		\left[{\delta_n \over \delta_m}\right]
		\left[ {I_{ul,m} \over I_{ul,n}} \right].
\end{equation}	
Aside from the, possibly large, deviation due to optical depth
effects, $\delta(\tau_{ul},\sigma_{\rm v})$, the abundance ratio of
the two states is well represented directly by the integrated
intensity.

An alternate and direct measure of the abundance ratio of these two
states can be found by taking the ratio of the column densities, via
equation \ref{e:Nm_2};
\begin{equation}
{N_{l,m} \over N_{l,n}} =\left[{1 - \exp(-T_{ul,n}/T_{{\rm ex},n}) \over 1 - \exp(-T_{ul,m}/T_{{\rm ex},m}) }\right]
				       \,\left[ {D_{ul,n}\over D_{ul,m}} \right]
				         \left[ {\tau_{ul,m} \, \sigma_{{\rm v},m} \over
				         \tau_{ul,n}  \,  \sigma_{{\rm v},n}} \right].
\end{equation}
Again assuming that $T_{{\rm ex}}\gg T_{ul}$ for both species and
$T_{{\rm ex},m}\approx T_{{\rm ex},n}$, this yields
\begin{equation}\label{e:ratio2}
{N_{l,m} \over N_{l,n}} =       \left[ {T_{ul,n}\, D_{ul,n} \over T_{ul,m} \, D_{ul,m}} \right]
				\left[ {\tau_{ul,m} \, \sigma_{{\rm v},m} \over \tau_{ul,n} \, \sigma_{{\rm v},n}} \right].			
\end{equation}

\subsubsection{Total Column Density Determination}
\label{sec:app:mol2}

The observational measures only probe the column density in a single
state $N_l$, which can be related to the total column density of the
species through the partition function $Z$.  Thus, if $Z_p$ denotes
the number of molecules in state $p$,
\begin{equation}
Z = \sum_p Z_p,
\end{equation}
and the conversion factor from $N_{l}$ to $N$ is 
\begin{equation}
N = \left({ Z \over Z_{l}} \right)\, N_{l}.
\end{equation}

For ammonia, there are two terms in the partition function to
consider: the distribution across the metastable rotational states of
the molecule and the distribution between the two levels of the
inversion transition within each rotational state.  Assuming LTE, the
partition function elements for the metastable rotational states ($J
= K$) are \citep{roh04};
\begin{equation}
Z_J = (2J + 1)\exp
       \left\{ -\frac{h}{kT}\,\left[ B\,J(J + 1) + (C-B)J^2 \right] \right\} ,
\end{equation}
where $B = 298117$~MHz and $C = 186726$~MHz are the rotational
constants of the ammonia molecule \citep{pic98}.  
Care must be taken, however, in calculating the total partition
function since the ortho- and para- species are not expected to
exchange. Thus, there are two separate partition functions, one for each of the
ortho- and para- states. Since we are only
treating para-ammonia (p-NH$_3$) in this paper, $J\neq 0,3,6,9$...  Additionally, the $J \neq K$
non-meta-stable states are assumed to carry no weight.  

The full partition function contains a combination of the partition function elements
for the appropriate metastable states (ortho or para) along with the partition function
elements of the inversion transition, for which the upper and lower states  
have equal statistical weight. Thus, the full partition function for p-NH$_3$ ($J = 1,2,4,5$...) is
\begin{equation}
\label{e:znh3}
Z_{\mathrm{p-NH3}} = \sum_J \left[1+\exp\left(-\frac{T_{ul}(J)}{T}\right)\right] (2J + 1)\exp
       \left\{ -\frac{h}{kT}\,\left[ B\,J(J + 1) + (C-B)J^2 \right] \right\},
\end{equation}
where the first term in square braces accounts for the two inversion levels within each rotational state.
For all inversion transitions, $T_{ul}(J)\sim 1$~K and therefore the first term in
the partition function $\approx 2$ for all relevant core temperatures. The observed p-\nhhh\ column density, however, corresponds only to the lower inversion level of the $J=1$ rotation state. When determining $Z_{1,\mathrm{p-NH3}}$
only the first term in the square brace should be included, hence
\begin{equation}
Z_{1,\mathrm{p-NH3}} =  3\,\exp \left[ -\frac{h(B+C)}{kT}\right].
\end{equation}
 Further, the temperature used in the
evaluation of the partition function, $T$ should characterize the
level distribution of the ammonia molecules.  Since transitions
between the metastable states are regulated by collisions, this is
often equated with the kinetic temperature of the gas $T_k$ or estimated from the observed level populations directly.



The \nnhp\ molecule is a linear rotator, so its partition function is
given by 
\begin{equation}
Z_{\mathrm{N2H+}} = \sum_J  (2J + 1)\exp
       \left\{ -\frac{h}{kT}\,\left[ B\,J(J + 1)\right] \right\}.
\end{equation}
where $B$ is the rotational constant: $B=46586.867$~MHz
\citep{pic98}.  In the case of \nnhp, $Z_0=1$ corresponding to the $J=0$ state.

\begin{figure} 
\epsscale{0.8}
\includegraphics[width=1.0\textwidth]{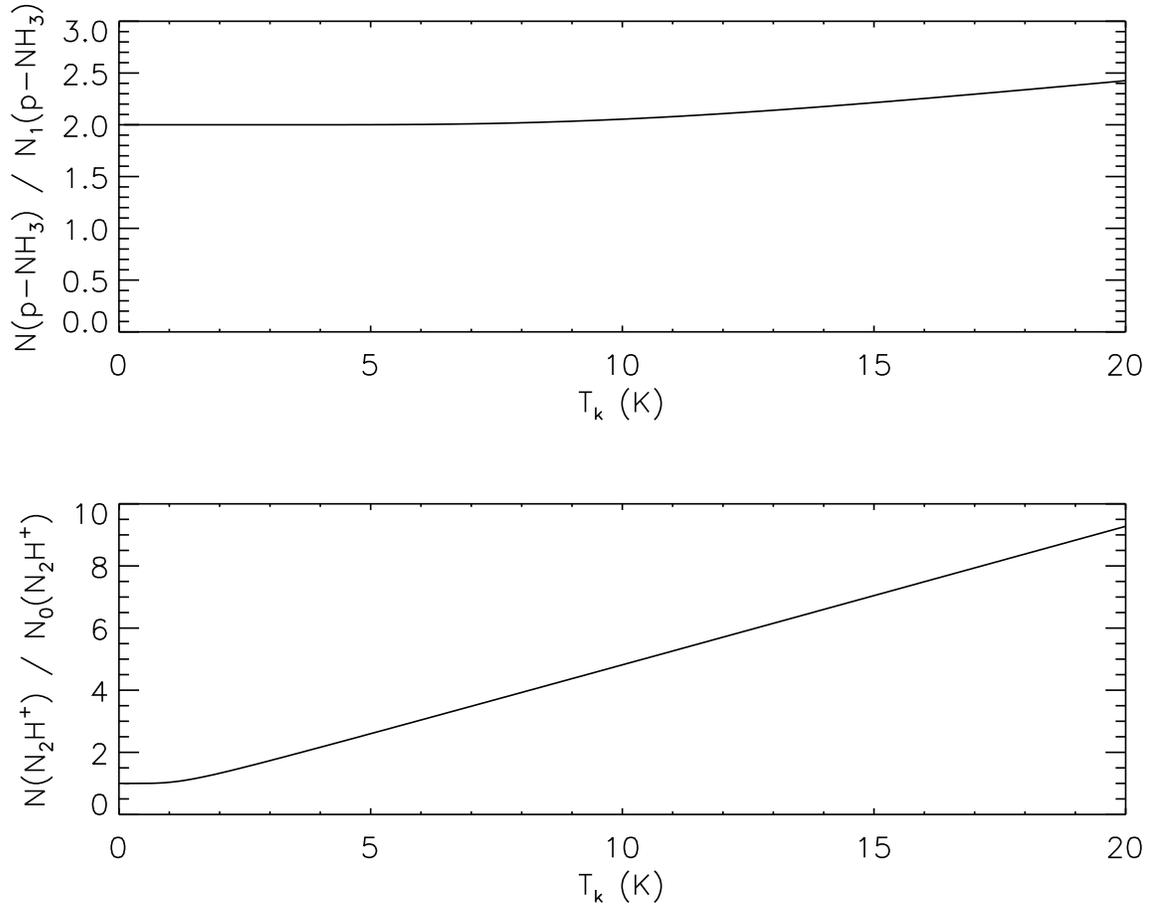}
\caption{Conversion factors required to convert the measured column density in an observed transition to the total column density of the species as a function of temperature (see text). The top panel shows the conversion required from $N_1$(p-\nhhh) [the lower state of the $(1,1)$ inversion level] to $N$(p-\nhhh). The bottom panel shows the conversion required from $N_0$(\nnhp) [the lower state of the (1-0) transition] to $N$(\nnhp).}
\label{fig:partition}
\end{figure}

Figure~\ref{fig:partition} plots the conversion factor ($N/N_l$, where $l = 1$ 
for p-\nhhh\ and $l = 0$ for \nnhp) as a function
of excitation temperature for both p-\nhhh\ and \nnhp. For an assumed
kinetic temperature of $T_k = 11\,$K, the required conversion factors
are 2.1 and 5.3, respectively. Careful consideration of
Figure~\ref{fig:partition} reveals that the conversion for
p-\nhhh\ (1,1) is quite insensitive to sub-thermal excitation of the
meta-stable states (i.e., deviation from LTE). The \nnhp\ (1--0)
conversion, however, is quite steep at these excitation
temperatures. For example, assuming that the effective temperature in
the partition function is 6.9\,K, similar to the measured excitation
temperatures for proto-stellar cores, yields a conversion for \nnhp\ of 3.4. 
An effective temperature of 5.7\,K, typical for pre-stellar cores in Perseus,
yields a conversion of only 2.9.

\end{appendix}

\newpage

%
%
\begin{deluxetable}{ccccccccccc}
\tablewidth{0pt}
\tablecolumns{11}
\tabletypesize{\tiny}
\rotate
\tablecaption{Perseus Core Cross-matched Observations\label{tab1}}
\tablehead{
\colhead{\nnhp \tablenotemark{a}} & 
\colhead{$\alpha_{\mathrm{J2000}}$\tablenotemark{a}} &
\colhead{$\delta_{\mathrm{J2000}}$\tablenotemark{a}} & 
\colhead{\nhhh \tablenotemark{b}} &
\colhead{Offset\tablenotemark{c}} & 
\colhead{Spitzer\tablenotemark{a}} &  
\colhead{$I$(\nhhh)\tablenotemark{b}} & 
\colhead{$I$(\nnhp)\tablenotemark{a}} & 
\colhead{$I$(\ceo)\tablenotemark{a}}& 
\colhead{$S_{850}$} & \colhead{$S_{1100}$} 
\\
\colhead{Src} & 
\colhead{($^\circ$)} & 
\colhead{($^\circ$)} & 
\colhead{Src} &
\colhead{($\arcsec$)} & 
\colhead{Src} &
\colhead{(K km s$^{-1}$)} & 
\colhead{(K km s$^{-1}$)} &
\colhead{(K km s$^{-1}$)} &
\colhead{(Jy bm$^{-1}$)\tablenotemark{d}} &
\colhead{(Jy bm$^{-1}$)\tablenotemark{d}}
}
\startdata
\cutinhead{Pre-Stellar Cores}
150 & $ 03:25:25.7$ & $ +30:45:02$ &  8    & 6.7 & \nodata & 12.6 & 9.4 & 2.9 & 1.21 & 0.35 \\
147 & $ 03:25:46.3$ & $ +30:44:14$ &  15   & 3.7 & \nodata & 5.8 & 4.3 & 3.7 & 0.24 & 0.20 \\
146 & $ 03:25:49.3$ & $ +30:42:15$ &  17   & 10.7 & \nodata & 11.7 & 6.2 & 3.1 & 0.99 & 0.37 \\
139 & $ 03:27:29.5$ & $ +30:15:09$ &  29   & 8.9 & \nodata & 4.6 & 2.6 & 0.8 & 0.27 & 0.10 \\
131 & $ 03:27:55.6$ & $ +30:06:05$ &  36   & 13.0 & \nodata & 5.1 & 1.8 & 0.4 & \nodata & \nodata \\
127 & $ 03:28:34.4$ & $ +30:19:25$ &  42   & 16.6 & \nodata & 2.3 & 2.3 & 1.1 & 0.15 & 0.11 \\
123 & $ 03:28:39.1$ & $ +31:18:24$ &  47   & 11.9 & \nodata & 12.9 & 9.1 & 3.6 & 1.05 & 0.31 \\
121 & $ 03:28:42.5$ & $ +31:06:13$ &  50   & 1.4 & \nodata & 8.4 & 6.0 & 2.4 & 0.52 & 0.16 \\
110 & $ 03:29:06.5$ & $ +31:15:36$ &  70   & 9.0 & \nodata & 13.2 & 16.6 & 8.3 & 2.00 & 0.58 \\
109 & $ 03:29:06.8$ & $ +31:17:18$ &  72   & 13.2 & \nodata & 7.5 & 6.5 & 6.3 & 0.96 & 0.45 \\
108 & $ 03:29:07.4$ & $ +31:21:48$ &  71   & 5.6 & \nodata & 0.8 & \nodata & 11.0 & 1.22 & 0.52 \\
107 & $ 03:29:08.8$ & $ +31:15:13$ &  73   & 1.4 & \nodata & 18.8 & 16.1 & 4.7 & 2.03 & 0.68 \\
105 & $ 03:29:10.2$ & $ +31:21:43$ &  76   & 1.9 & \nodata & 0.8 & \nodata & 9.3 & 1.38 & 0.55 \\
101 & $ 03:29:15.0$ & $ +31:20:32$ &  79   & 5.3 & \nodata & 2.1 & 2.5 & 10.7 & 0.59 & 0.30 \\
99 & $ 03:29:18.4$ & $ +31:25:03$ & 82.2  & 10.7 & \nodata & 4.7 & 3.4 & 4.0 & 0.99 & 0.30 \\
96 & $ 03:29:25.1$ & $ +31:28:16$ & 88.1  & 8.7 & \nodata & 7.8 & 5.6 & 1.8 & 0.52 & 0.20 \\
94 & $ 03:30:15.0$ & $ +30:23:45$ &  91   & 6.5 & \nodata & 7.6 & 4.7 & 1.6 & 0.35 & 0.11 \\
92 & $ 03:30:32.0$ & $ +30:26:24$ &  95   & 5.2 & \nodata & 11.1 & 6.0 & 1.1 & 0.25 & 0.18 \\
91 & $ 03:30:46.1$ & $ +30:52:44$ &  96   & 10.9 & \nodata & 2.6 & 3.2 & 1.8 & 0.27 & 0.13 \\
85 & $ 03:32:27.4$ & $ +30:59:22$ &  104  & 12.7 & \nodata & 6.7 & 3.4 & 2.7 & 0.47 & 0.15 \\
84 & $ 03:32:28.6$ & $ +31:02:10$ &  105  & 10.6 & \nodata & 3.9 & 3.1 & 2.1 & 0.28 & 0.09 \\
80 & $ 03:32:38.2$ & $ +30:57:28$ &  107  & 14.4 & \nodata & 1.5 & 0.9 & 2.4 & 0.19 & 0.13 \\
79 & $ 03:32:43.2$ & $ +31:00:00$ &  108  & 12.2 & \nodata & 8.9 & 4.0 & 3.8 & 0.48 & 0.20 \\
78 & $ 03:32:50.6$ & $ +31:01:49$ &  109  & 9.7 & \nodata & 1.8 & 1.4 & 3.2 & 0.30 & 0.17 \\
77 & $ 03:32:58.0$ & $ +31:03:19$ &  111  & 12.7 & \nodata & 6.0 & 3.4 & 2.6 & 0.48 & 0.22 \\
76 & $ 03:33:00.6$ & $ +31:20:50$ &  112  & 8.5 & \nodata & 1.8 & 1.6 & 3.1 & 0.29 & 0.16 \\
75 & $ 03:33:04.0$ & $ +31:04:57$ &  114  & 4.0 & \nodata & 12.1 & 5.9 & 3.3 & 0.66 & 0.22 \\
67 & $ 03:33:31.2$ & $ +31:20:11$ &  127  & 12.5 & \nodata & 3.6 & 1.8 & 1.8 & 0.21 & 0.12 \\
63 & $ 03:35:20.8$ & $ +31:07:05$ &  132  & 12.4 & \nodata & 0.7 & 0.3 & 1.6 & 0.17 & 0.08 \\
41 & $ 03:40:49.7$ & $ +31:48:34$ &  142  & 3.1 & \nodata & 2.2 & 1.5 & 2.8 & 0.29 & 0.13 \\
34 & $ 03:41:44.2$ & $ +31:48:14$ &  146  & 12.8 & \nodata & 0.6 & \nodata & 3.3 & 0.17 & 0.10 \\
36 & $ 03:41:45.8$ & $ +31:57:22$ &  147  & 2.9 & \nodata & 5.9 & 2.8 & 1.1 & 0.25 & 0.09 \\
33 & $ 03:41:59.1$ & $ +31:58:27$ &  148  & 13.7 & \nodata & 1.3 & 0.6 & 1.8 & 0.15 & 0.08 \\
30 & $ 03:42:48.1$ & $ +31:58:51$ &  152  & 15.2 & \nodata & 1.0 & 0.8 & 3.5 & 0.32 & 0.13 \\
28 & $ 03:43:38.3$ & $ +32:03:06$ &  156  & 4.5 & \nodata & 4.4 & 3.3 & 4.6 & 0.54 & 0.16 \\
27 & $ 03:43:44.0$ & $ +32:02:46$ & 157.2 & 5.7 & \nodata & 4.0 & 3.2 & 4.5 & 0.69 & 0.22 \\
23 & $ 03:43:58.2$ & $ +32:04:01$ &  163  & 6.7 & \nodata & 3.8 & 4.0 & 6.3 & 0.71 & 0.27 \\
19 & $ 03:44:06.6$ & $ +32:02:06$ &  169  & 16.6 & \nodata & 5.0 & 4.8 & 5.0 & 0.68 & 0.25 \\
18 & $ 03:44:36.9$ & $ +31:58:41$ &  176  & 6.2 & \nodata & 2.9 & 3.0 & 3.8 & 0.53 & 0.15 \\
15 & $ 03:44:48.9$ & $ +32:00:32$ &  180  & 2.5 & \nodata & 3.3 & 3.0 & 3.8 & 0.49 & 0.13 \\
6 & $ 03:47:39.0$ & $ +32:52:11$ &  189  & 8.9 & \nodata & 6.2 & 4.4 & 3.2 & 0.63 & 0.24 \\
5 & $ 03:47:40.4$ & $ +32:54:14$ &  190  & 18.8 & \nodata & 1.9 & 1.6 & 2.6 & 0.08 & 0.08 \\

\cutinhead{Proto-Stellar Cores}
152 & $ 03:25:22.5$ & $ +30:45:07$ &  7    & 4.1 &            1 & 10.5 & 10.8 & 4.2 & 2.04 & 0.66 \\
149 & $ 03:25:36.0$ & $ +30:45:11$ &  12   & 2.8 &            2 & 20.3 & 20.2 & 7.4 & 7.00 & 1.98 \\
148 & $ 03:25:38.9$ & $ +30:44:00$ &  14   & 4.4 &            3 & 13.6 & 13.0 & 4.0 & 2.69 & 0.87 \\
143 & $ 03:26:37.2$ & $ +30:15:19$ &  22   & 4.4 &            4 & 5.0 & 4.3 & 0.9 & 0.41 & 0.18 \\
136 & $ 03:27:37.9$ & $ +30:13:53$ & 31.1  & 6.7 &            5 & 6.2 & 6.6 & 0.9 & 0.50 & 0.15 \\
135 & $ 03:27:39.0$ & $ +30:12:54$ &  32   & 6.5 &            6 & 10.8 & 9.4 & 1.1 & 1.12 & 0.27 \\
134 & $ 03:27:42.7$ & $ +30:12:24$ &  34   & 11.3 &            7 & 11.0 & 9.5 & 1.7 & 0.89 & 0.26 \\
133 & $ 03:27:48.3$ & $ +30:12:08$ &  35   & 7.5 &            8 & 6.6 & 6.8 & 1.2 & 0.78 & 0.21 \\
128 & $ 03:28:32.3$ & $ +31:10:59$ &  40   & 10.8 &            9 & 5.5 & 4.8 & 1.7 & 0.40 & 0.09 \\
126 & $ 03:28:34.5$ & $ +31:06:59$ &  43   & 5.8 &           10 & 3.1 & 3.5 & 1.0 & 0.21 & 0.12 \\
125 & $ 03:28:36.7$ & $ +31:13:24$ &  44   & 6.2 &           11 & 5.4 & 5.8 & 3.0 & 1.14 & 0.25 \\
124 & $ 03:28:38.8$ & $ +31:05:54$ &  46   & 6.4 &           12 & 9.0 & 6.3 & 3.8 & 0.60 & 0.18 \\
122 & $ 03:28:40.1$ & $ +31:17:48$ &  48   & 7.7 &           13 & 16.5 & 10.3 & 2.5 & 1.14 & 0.36 \\
118 & $ 03:28:55.3$ & $ +31:14:28$ &  58   & 5.4 &           15 & 11.4 & 12.4 & 7.3 & 4.40 & 1.10 \\
116 & $ 03:28:59.5$ & $ +31:21:29$ &  64   & 9.1 &           17 & 8.7 & 9.9 & 8.8 & 1.82 & 0.64 \\
115 & $ 03:29:00.2$ & $ +31:11:53$ &  65   & 8.0 &           18 & 8.1 & 8.5 & 3.3 & 0.38 & 0.17 \\
113 & $ 03:29:01.4$ & $ +31:20:23$ &  66   & 11.3 &           19 & 11.7 & 11.8 & 13.7 & 3.07 & 1.02 \\
112 & $ 03:29:03.2$ & $ +31:15:54$ &  67   & 5.4 &           20 & 13.2 & 18.5 & 10.3 & 6.42 & 2.01 \\
111 & $ 03:29:03.7$ & $ +31:14:48$ &  68   & 11.4 &           21 & 17.1 & 17.6 & 5.1 & 1.89 & 0.84 \\
106 & $ 03:29:09.9$ & $ +31:13:31$ &  75   & 6.5 &           22 & 11.4 & 11.4 & 7.0 & 10.58 & 4.63 \\
104 & $ 03:29:10.5$ & $ +31:18:25$ &  77   & 11.3 &           23 & 10.3 & 10.4 & 7.0 & 2.71 & 0.78 \\
103 & $ 03:29:11.3$ & $ +31:13:07$ &  78   & 1.5 &           25 & 8.0 & 7.9 & 2.0 & 8.44 & 2.62 \\
100 & $ 03:29:17.2$ & $ +31:27:44$ &  81   & 4.4 &           27 & 4.2 & 5.3 & 5.3 & 0.60 & 0.26 \\
98 & $ 03:29:18.5$ & $ +31:23:09$ &  84   & 21.1 &           28 & 2.5 & 2.7 & 6.7 & 0.76 & 0.31 \\
97 & $ 03:29:23.4$ & $ +31:33:16$ &  87   & 6.1 &           30 & 7.0 & 5.9 & 2.0 & 0.53 & 0.20 \\
95 & $ 03:29:52.0$ & $ +31:39:03$ &  89   & 10.5 &           31 & 6.9 & 6.3 & 2.4 & 0.62 & 0.21 \\
90 & $ 03:31:19.1$ & $ +30:45:26$ &  99   & 13.1 &           32 & 9.0 & 4.6 & 2.0 & 1.03 & 0.46 \\
86 & $ 03:32:18.0$ & $ +30:49:45$ &  103  & 7.4 &           33 & 11.9 & 9.6 & 3.8 & 2.39 & 1.01 \\
74 & $ 03:33:13.8$ & $ +31:19:51$ &  118  & 6.2 &           34 & 10.0 & 4.5 & 2.1 & 0.62 & 0.25 \\
73 & $ 03:33:16.1$ & $ +31:06:52$ &  119  & 17.3 &           36 & 16.1 & 10.7 & 5.5 & 1.96 & 0.60 \\
72 & $ 03:33:17.9$ & $ +31:09:28$ &  121  & 2.2 &           38 & 16.8 & 13.4 & 7.2 & 3.23 & 0.95 \\
71 & $ 03:33:21.7$ & $ +31:07:22$ &  123  & 21.1 &           39 & 19.2 & 14.1 & 4.2 & 2.75 & 1.09 \\
68 & $ 03:33:27.3$ & $ +31:06:59$ &  126  & 3.6 &           40 & 4.4 & 4.2 & 5.9 & 0.43 & 0.17 \\
26 & $ 03:43:51.1$ & $ +32:03:21$ &  160  & 8.5 &           41 & 8.4 & 7.3 & 2.9 & 1.09 & 0.40 \\
25 & $ 03:43:57.0$ & $ +32:00:50$ &  161  & 13.7 &           44 & 9.7 & 6.9 & 6.5 & 2.99 & 0.95 \\
24 & $ 03:43:57.2$ & $ +32:03:02$ &  162  & 2.9 &           43 & 3.7 & 4.1 & 5.4 & 2.18 & 0.72 \\
21 & $ 03:44:01.8$ & $ +32:01:55$ &  164  & 7.6 &           47 & 4.5 & 6.4 & 5.3 & 1.01 & 0.37 \\
20 & $ 03:44:03.3$ & $ +32:02:24$ &  165  & 16.0 &           47 & 6.9 & 5.5 & 4.0 & 0.84 & 0.36 \\
16 & $ 03:44:44.2$ & $ +32:01:27$ &  178  & 3.8 &           48 & 1.7 & 3.1 & 5.9 & 1.14 & 0.28 \\
4 & $ 03:47:41.8$ & $ +32:51:40$ &  192  & 9.3 &           49 & 6.9 & 7.4 & 3.9 & 0.81 & 0.24 \\

\enddata
\tablenotetext{a}{From \citet{kir07}. }
\tablenotetext{b}{From \citet{ros08}.}
\tablenotetext{c}{Separation between \citet{kir07} and \citet{ros08} pointings.}
\tablenotetext{d}{Within a $30''$ beam.}
\end{deluxetable}

%
%
\begin{deluxetable}{lrrrrrrrrrrrrrr}
\tablecaption{Line Properties \label{tab2}}
\tabletypesize{\scriptsize}
\rotate
\tablecolumns{15}
\tablehead{\colhead{} & \multicolumn{5}{c}{NH$_3$ Properties\tablenotemark{a}} & &
\multicolumn{4}{c}{N$_2$H$^+$ Properties\tablenotemark{b}} & &
\multicolumn{3}{c}{C$^{18}$O Properties\tablenotemark{b}} \\
\cline{2-6} \cline{8-11} \cline{13-14} \\
\colhead{NH$_3$\tablenotemark{a}} & 
\colhead{$V_{\mathrm{LSR}}$} & \colhead{$\sigma_{\mathrm{v}}$} &
\colhead{$\tau$} &  \colhead{$T_{\mathrm{k}}$} &\colhead{$T_{\mathrm{ex}}$} & &
\colhead{$V_{\mathrm{LSR}}$} & \colhead{$\sigma_{\mathrm{v}}$} &
\colhead{$\tau$} & \colhead{$T_{\mathrm{ex}}$} & &
\colhead{$V_{\mathrm{LSR}}$} & \colhead{$\sigma_{\mathrm{v}}$} \\
\colhead{Src} & \colhead{(km s$^{-1}$)} &
 \colhead{(km s$^{-1}$)} & & \colhead{(K)} & \colhead{(K)} && 
 \colhead{(km s$^{-1}$)} &
 \colhead{(km s$^{-1}$)} & &\colhead{(K)} & &
 \colhead{(km s$^{-1}$)} & \colhead{(km s$^{-1}$)}
}
\startdata
\cutinhead{Pre-Stellar Cores}
 8    & 4.09 & 0.16 & 10.9 & 11.2 & 7.7 &  & 4.05 & 0.21 & 13.0 & 5.7 &  & 4.09 & 0.29 \\
 15   & 4.65 & 0.19 & 3.7 & 11.2 & 6.7 &  & 4.64 & 0.20 & 4.6 & 5.4 &  & 4.55 & 0.37 \\
 17   & 4.51 & 0.15 & 14.7 & 9.1 & 6.8 &  & 4.49 & 0.16 & 9.1 & 5.6 &  & 4.34 & 0.37 \\
 29   & 5.08 & 0.13 & 4.7 & 10.6 & 6.5 &  & 5.09 & 0.17 & 2.5 & 5.7 &  & 4.92 & 0.41 \\
 36   & 4.71 & 0.08 & 9.7 & 9.1 & 6.4 &  & 4.72 & 0.08 & 3.7 & 6.0 &  & 4.89 & 0.33 \\
 42   & 5.47 & 0.14 & 3.8 & 10.4 & 4.9 &  & 5.51 & 0.16 & 1.0 & 9.2 &  & 5.52 & 0.20 \\
 47   & 8.18 & 0.19 & 8.2 & 11.7 & 7.9 &  & 8.19 & 0.20 & 10.6 & 6.0 &  & 8.34 & 0.40 \\
 50   & 7.21 & 0.16 & 7.9 & 10.5 & 6.8 &  & 7.22 & 0.16 & 6.2 & 6.4 &  & 7.28 & 0.39 \\
 70   & 8.02 & 0.36 & 3.1 & 14.9 & 8.7 &  & 7.97 & 0.39 & 2.4 & 11.9 &  & 7.56 & 0.80 \\
 72   & 8.48 & 0.21 & 3.3 & 12.6 & 7.8 &  & 8.48 & 0.23 & 3.0 & 7.6 &  & 8.52 & 0.63 \\
 73   & 7.76 & 0.46 & 4.1 & 12.3 & 8.4 &  & 7.93 & 0.32 & 5.1 & 7.4 &  & 7.75 & 0.68 \\
82.2  & 7.51 & 0.11 & 3.7 & 13.4 & 5.4 &  & 7.54 & 0.07 & 10.8 & 3.1 &  & 7.53 & 0.17 \\
88.1  & 7.41 & 0.13 & 6.4 & 10.3 & 6.1 &  & 7.54 & 0.21 & 7.3 & 5.3 &  & 7.53 & 0.30 \\
 91   & 5.88 & 0.14 & 7.5 & 11.0 & 6.9 &  & 5.90 & 0.16 & 2.5 & 8.2 &  & 5.98 & 0.28 \\
 95   & 6.06 & 0.15 & 13.8 & 10.0 & 6.8 &  & 6.08 & 0.16 & 14.0 & 5.1 &  & 6.06 & 0.24 \\
 96   & 7.81 & 0.15 & 2.9 & 10.5 & 5.5 &  & 7.83 & 0.16 & 8.2 & 4.4 &  & 7.78 & 0.24 \\
 104  & 6.42 & 0.15 & 7.7 & 10.5 & 6.1 &  & 6.41 & 0.17 & 14.6 & 4.0 &  & 6.50 & 0.42 \\
 105  & 6.65 & 0.11 & 7.6 & 9.7 & 5.2 &  & 6.65 & 0.12 & 4.6 & 5.8 &  & 6.77 & 0.29 \\
 108  & 6.80 & 0.13 & 11.0 & 10.2 & 6.7 &  & 6.82 & 0.16 & 9.6 & 4.6 &  & 6.57 & 0.43 \\
 109  & 6.58 & 0.23 & 1.3 & 11.2 & 5.2 &  & 6.57 & 0.18 & 2.7 & 4.2 &  & 6.65 & 0.32 \\
 111  & 6.67 & 0.14 & 6.2 & 10.1 & 6.5 &  & 6.65 & 0.13 & 10.6 & 4.6 &  & 6.70 & 0.24 \\
 112  & 6.58 & 0.26 & 1.1 & 11.5 & 5.5 &  & 6.51 & 0.21 & 9.1 & 3.4 &  & 6.63 & 0.31 \\
 114  & 6.61 & 0.19 & 9.9 & 9.9 & 7.1 &  & 6.59 & 0.18 & 13.5 & 4.9 &  & 6.68 & 0.34 \\
 127  & 6.30 & 0.26 & 2.1 & 11.0 & 5.4 &  & 6.22 & 0.15 & 1.2 & 7.1 &  & 6.29 & 0.23 \\
 142  & 8.47 & 0.11 & 3.0 & 12.2 & 5.2 &  & 8.44 & 0.10 & 4.7 & 4.7 &  & 8.39 & 0.16 \\
 147  & 9.43 & 0.10 & 9.6 & 9.7 & 6.6 &  & 9.41 & 0.09 & 6.7 & 5.7 &  & 9.35 & 0.14 \\
 148  & 9.43 & 0.15 & 3.6 & 17.9 & 4.1 &  & 9.40 & 0.15 & 1.2 & 5.1 &  & 9.29 & 0.24 \\
 152  & 8.82 & 0.14 & 1.7 & 14.0 & 4.7 &  & 8.84 & 0.09 & 2.9 & 4.5 &  & 8.74 & 0.19 \\
 156  & 8.55 & 0.35 & 1.7 & 13.5 & 6.1 &  & 8.46 & 0.25 & 6.1 & 4.3 &  & 8.67 & 0.42 \\
157.2 & 8.74 & 0.14 & 3.0 & 12.3 & 6.6 &  & 8.70 & 0.12 & 3.0 & 5.2 &  & 8.72 & 0.24 \\
 169  & 8.48 & 0.16 & 4.2 & 11.4 & 6.5 &  & 8.55 & 0.15 & 4.1 & 7.1 &  & 8.56 & 0.29 \\
 176  & 9.97 & 0.17 & 2.5 & 10.8 & 5.8 &  & 9.92 & 0.22 & 4.0 & 4.7 &  & 9.59 & 0.52 \\
 180  & 8.95 & 0.12 & 4.3 & 10.8 & 5.8 &  & 8.94 & 0.12 & 4.9 & 5.4 &  & 9.10 & 0.36 \\
 189  & 10.37 & 0.18 & 5.1 & 10.1 & 6.2 &  & 10.37 & 0.18 & 6.2 & 5.3 &  & 10.18 & 0.36 \\
 190  & 10.13 & 0.12 & 3.0 & 10.1 & 5.2 &  & 10.10 & 0.13 & 2.9 & 5.3 &  & 10.06 & 0.34 \\

\cutinhead{Proto-Stellar Cores}
 7    & 4.11 & 0.18 & 7.0 & 12.1 & 7.6 &  & 4.08 & 0.21 & 12.8 & 6.1 &  & 4.05 & 0.31 \\
 12   & 4.53 & 0.38 & 5.1 & 12.6 & 8.7 &  & 4.51 & 0.35 & 6.4 & 8.8 &  & 4.35 & 0.45 \\
 22   & 5.15 & 0.15 & 6.3 & 11.7 & 5.7 &  & 5.09 & 0.12 & 8.9 & 5.4 &  & 5.09 & 0.19 \\
31.1  & 4.64 & 0.17 & 3.7 & 11.7 & 5.8 &  & 4.64 & 0.24 & 1.9 & 8.2 &  & 5.90 & 0.19 \\
 32   & 4.76 & 0.27 & 4.4 & 13.1 & 7.6 &  & 4.72 & 0.26 & 6.6 & 6.3 &  & 4.68 & 0.30 \\
 34   & 4.97 & 0.17 & 6.1 & 11.4 & 8.4 &  & 4.87 & 0.20 & 6.0 & 7.4 &  & 4.91 & 0.30 \\
 35   & 4.95 & 0.21 & 3.8 & 11.9 & 6.8 &  & 5.06 & 0.23 & 4.6 & 6.3 &  & 5.22 & 0.45 \\
 40   & 7.19 & 0.15 & 6.9 & 10.8 & 5.7 &  & 7.21 & 0.16 & 6.7 & 5.6 &  & 7.14 & 0.59 \\
 43   & 6.82 & 0.11 & 6.1 & 10.6 & 5.2 &  & 6.84 & 0.13 & 3.4 & 6.5 &  & 6.86 & 0.70 \\
 46   & 7.03 & 0.15 & 7.0 & 10.5 & 7.4 &  & 7.04 & 0.15 & 7.1 & 6.6 &  & 6.91 & 0.66 \\
 48   & 7.99 & 0.19 & 8.8 & 11.7 & 9.1 &  & 7.98 & 0.23 & 7.5 & 6.8 &  & 8.06 & 0.41 \\
 58   & 7.48 & 0.37 & 2.5 & 16.5 & 8.5 &  & 7.34 & 0.24 & 4.3 & 6.6 &  & 7.62 & 0.56 \\
 64   & 7.79 & 0.31 & 2.2 & 14.4 & 8.7 &  & 7.83 & 0.33 & 2.2 & 9.6 &  & 8.00 & 0.65 \\
 65   & 7.15 & 0.21 & 4.5 & 12.5 & 7.1 &  & 7.16 & 0.27 & 3.5 & 7.6 &  & 7.46 & 0.65 \\
 66   & 8.01 & 0.27 & 2.8 & 16.4 & 10.1 &  & 7.97 & 0.25 & 4.9 & 8.3 &  & 7.73 & 0.62 \\
 67   & 8.44 & 0.30 & 3.6 & 16.3 & 8.8 &  & 8.40 & 0.29 & 4.7 & 10.6 &  & 8.18 & 0.64 \\
 68   & 7.43 & 0.58 & 2.3 & 16.4 & 9.2 &  & 7.74 & 0.36 & 5.1 & 5.7 &  & 7.39 & 0.91 \\
 75   & 7.32 & 0.60 & 2.1 & 15.1 & 7.2 &  & 7.49 & 0.41 & 5.6 & 4.9 &  & 7.30 & 0.82 \\
 77   & 8.57 & 0.22 & 3.5 & 14.3 & 9.3 &  & 8.59 & 0.23 & 4.3 & 8.4 &  & 8.54 & 0.60 \\
 78   & 7.13 & 0.54 & 1.9 & 14.3 & 6.5 &  & 7.15 & 0.53 & 2.3 & 5.2 &  & 6.60 & 0.36 \\
 81   & 7.49 & 0.13 & 4.2 & 11.8 & 6.5 &  & 7.52 & 0.14 & 4.4 & 7.5 &  & 7.55 & 0.30 \\
 84   & 7.48 & 0.15 & 3.4 & 13.4 & 5.1 &  & 7.47 & 0.17 & 3.8 & 5.1 &  & 7.39 & 0.44 \\
 87   & 7.49 & 0.13 & 6.6 & 10.4 & 7.4 &  & 7.50 & 0.13 & 7.3 & 6.8 &  & 7.51 & 0.21 \\
 89   & 8.18 & 0.13 & 5.5 & 10.5 & 8.0 &  & 8.15 & 0.15 & 7.3 & 6.3 &  & 8.10 & 0.22 \\
 99   & 6.99 & 0.18 & 6.0 & 11.4 & 7.4 &  & 7.01 & 0.25 & 2.1 & 7.1 &  & 7.08 & 0.67 \\
 103  & 6.88 & 0.24 & 5.8 & 11.6 & 7.7 &  & 6.90 & 0.25 & 5.3 & 7.1 &  & 6.89 & 0.41 \\
 118  & 6.83 & 0.13 & 15.9 & 9.4 & 6.4 &  & 6.84 & 0.14 & 9.1 & 5.3 &  & 6.93 & 0.37 \\
 119  & 6.41 & 0.26 & 8.1 & 11.5 & 7.8 &  & 6.43 & 0.24 & 13.0 & 5.8 &  & 6.65 & 0.40 \\
 121  & 6.25 & 0.33 & 5.4 & 12.4 & 8.3 &  & 6.27 & 0.32 & 5.9 & 7.3 &  & 6.47 & 0.51 \\
 123  & 6.60 & 0.33 & 7.2 & 11.7 & 7.8 &  & 6.66 & 0.37 & 8.1 & 6.5 &  & 6.41 & 0.36 \\
 160  & 8.64 & 0.19 & 3.5 & 11.7 & 8.7 &  & 8.56 & 0.18 & 4.2 & 8.0 &  & 8.52 & 0.24 \\
 161  & 8.98 & 0.22 & 4.3 & 12.9 & 7.7 &  & 9.06 & 0.20 & 5.1 & 6.6 &  & 8.94 & 0.34 \\
 162  & 8.74 & 0.22 & 1.7 & 14.2 & 7.3 &  & 8.74 & 0.20 & 3.0 & 6.3 &  & 8.76 & 0.36 \\
 164  & 9.00 & 0.27 & 1.5 & 13.0 & 7.8 &  & 9.02 & 0.28 & 2.0 & 8.3 &  & 8.82 & 0.42 \\
 165  & 8.47 & 0.18 & 5.5 & 11.1 & 6.3 &  & 8.57 & 0.29 & 2.2 & 6.9 &  & 8.55 & 0.26 \\
 192  & 10.24 & 0.16 & 6.0 & 11.7 & 6.8 &  & 10.24 & 0.19 & 6.3 & 6.5 &  & 10.10 & 0.39 \\

\enddata
\tablenotetext{a}{From \citet{ros08}.}
\tablenotetext{b}{From \citet{kir07}. }
\end{deluxetable}

%
%
\begin{deluxetable}{lrrr}
\tablewidth{0pt}
\tablecolumns{4}
\tablecaption{\label{tab3}Derived Column Densities for Dense Gas Tracers}
\tablehead{
\colhead{NH$_3$\tablenotemark{a}} & \colhead{$N$(p-\nhhh)} &
\colhead{$N$(\nnhp)} & \colhead{$N$(\hh)} \\
\colhead{Src} & \colhead{ ($10^{12}$~cm$^{-2}$)} &
\colhead{ ($10^{12}$~cm$^{-2}$)} & \colhead{ ($10^{20}$~cm$^{-2}$)} 
}
\startdata
\cutinhead{Pre-Stellar Cores}
 8    & 453 & 25.3 & 363 \\
 15   & 167 & 8.3 & 72 \\
 17   & 506 & 13.2 & 296 \\
 29   & 135 & 3.9 & 81 \\
 36   & 181 & 2.9 & \nodata \\
 42   & 91 & 2.0 & 45 \\
 47   & 421 & 20.8 & 314 \\
 50   & 293 & 10.0 & 155 \\
 70   & 331 & 15.0 & 599 \\
 72   & 186 & 7.7 & 287 \\
 73   & 547 & 18.0 & 609 \\
82.2  & 81 & 5.3 & 296 \\
88.1  & 186 & 13.5 & 157 \\
 91   & 256 & 4.9 & 106 \\
 95   & 478 & 19.3 & 76 \\
 96   & 86 & 10.2 & 81 \\
 104  & 253 & 19.0 & 141 \\
 105  & 157 & 5.1 & 83 \\
 108  & 332 & 12.6 & 143 \\
 109  & 55 & 3.9 & 91 \\
 111  & 197 & 11.5 & 143 \\
 112  & 55 & 13.0 & 88 \\
 114  & 466 & 20.3 & 197 \\
 127  & 106 & 1.9 & 64 \\
 142  & 63 & 3.8 & 88 \\
 147  & 217 & 5.9 & 76 \\
 148  & 81 & 1.5 & 46 \\
 152  & 40 & 2.2 & 95 \\
 156  & 130 & 11.7 & 162 \\
157.2 & 93 & 3.0 & 206 \\
 169  & 150 & 6.6 & 203 \\
 176  & 86 & 7.2 & 160 \\
 180  & 106 & 5.5 & 147 \\
 189  & 203 & 9.7 & 189 \\
 190  & 66 & 3.3 & 23 \\

\cutinhead{Proto-Stellar Cores}
 7    & 324 & 31.3 & 613 \\
 12   & 573 & 33.2 & 2100 \\
 22   & 186 & 11.4 & 124 \\
31.1  & 132 & 6.4 & 151 \\
 32   & 316 & 20.6 & 336 \\
 34   & 298 & 15.8 & 267 \\
 35   & 187 & 12.8 & 233 \\
 40   & 210 & 11.7 & 120 \\
 43   & 125 & 5.4 & 62 \\
 46   & 276 & 12.4 & 179 \\
 48   & 524 & 21.0 & 342 \\
 58   & 272 & 12.6 & 1321 \\
 64   & 199 & 11.3 & 546 \\
 65   & 232 & 12.8 & 114 \\
 66   & 260 & 17.6 & 922 \\
 67   & 326 & 23.3 & 1925 \\
 68   & 413 & 20.2 & 567 \\
 75   & 318 & 22.9 & 3175 \\
 77   & 244 & 14.2 & 812 \\
 78   & 228 & 12.4 & 2532 \\
 81   & 125 & 8.0 & 180 \\
 84   & 89 & 6.4 & 227 \\
 87   & 213 & 11.6 & 158 \\
 89   & 189 & 12.6 & 187 \\
 99   & 276 & 6.6 & 308 \\
 103  & 372 & 17.1 & 718 \\
 118  & 464 & 12.8 & 186 \\
 119  & 565 & 34.3 & 589 \\
 121  & 506 & 24.8 & 968 \\
 123  & 650 & 35.5 & 824 \\
 160  & 206 & 10.5 & 328 \\
 161  & 259 & 12.5 & 896 \\
 162  & 90 & 7.0 & 655 \\
 164  & 108 & 8.1 & 302 \\
 165  & 223 & 7.9 & 253 \\
 192  & 236 & 14.5 & 244 \\

\enddata
\tablenotetext{a}{From \citet{ros08}.}
\end{deluxetable}


\newpage

%
%

%
%

%
%

%
%

%
%

%
%

%
%

%
%

\end{document}